\pdfoutput=1

\documentclass[12pt,a4paper]{article}

\usepackage{ifthen} 
\usepackage{caption}
\usepackage{graphicx, subcaption}
\usepackage[percent]{overpic}
\newboolean{pdflatex}
\setboolean{pdflatex}{true} 

\newboolean{articletitles}
\setboolean{articletitles}{true} 

\newboolean{uprightparticles}
\setboolean{uprightparticles}{false} 
\usepackage{multirow}
\usepackage{comment}

\def\paperauthors{LHCb collaboration} 
\def\paperasciititle{Observation of excited $\Omega_c^0$ baryons in  $\Omega_b^- \to \Xi_c^+ K^-m\pi^-$ decays} 
\def\papertitle{Observation of excited \Omegac baryons in  \decay{\Omegab}{\Xicp\Km\pim} decays} 
\def\paperkeywords{{High Energy Physics}, {LHCb}} 
\def\papercopyright{\the\year\ CERN for the benefit of the LHCb collaboration} 
\def\paperlicence{CC BY 4.0 licence}
\def\paperlicenceurl{https://creativecommons.org/licenses/by/4.0/}

\usepackage[top=1in, bottom=1.25in, left=1in, right=1in]{geometry}

\usepackage{microtype}
\usepackage{lineno}  
\usepackage{xspace} 
\usepackage{caption} 

\usepackage{graphicx}  
\usepackage{color}
\usepackage{colortbl}
\graphicspath{{./figs/}} 

\usepackage{amsmath} 
\usepackage{amssymb}
\usepackage{amsfonts}
\usepackage{upgreek} 

\newcommand*\patchAmsMathEnvironmentForLineno[1]{%
\expandafter\let\csname old#1\expandafter\endcsname\csname #1\endcsname
\expandafter\let\csname oldend#1\expandafter\endcsname\csname
end#1\endcsname
 \renewenvironment{#1}%
   {\linenomath\csname old#1\endcsname}%
   {\csname oldend#1\endcsname\endlinenomath}%
}
\newcommand*\patchBothAmsMathEnvironmentsForLineno[1]{%
  \patchAmsMathEnvironmentForLineno{#1}%
  \patchAmsMathEnvironmentForLineno{#1*}%
}
\AtBeginDocument{%
\patchBothAmsMathEnvironmentsForLineno{equation}%
\patchBothAmsMathEnvironmentsForLineno{align}%
\patchBothAmsMathEnvironmentsForLineno{flalign}%
\patchBothAmsMathEnvironmentsForLineno{alignat}%
\patchBothAmsMathEnvironmentsForLineno{gather}%
\patchBothAmsMathEnvironmentsForLineno{multline}%
\patchBothAmsMathEnvironmentsForLineno{eqnarray}%
}

\usepackage{hyperxmp}

\usepackage[pdftex,
            pdfauthor={\paperauthors},
            pdftitle={\paperasciititle},
            pdfkeywords={\paperkeywords},
            pdfcopyright={Copyright (C) \papercopyright},
            pdflicenseurl={\paperlicenceurl}]{hyperref}

\usepackage[colorinlistoftodos,textsize=scriptsize]{todonotes}

\usepackage[bottom,flushmargin,hang,multiple]{footmisc}

\usepackage[all]{hypcap} 

\usepackage{xspace} 
\usepackage{upgreek}

\def\lhcb   {\mbox{LHCb}\xspace}

\def\MagUp {\mbox{\em Mag\kern -0.05em Up}\xspace}

\ifthenelse{\boolean{uprightparticles}}%
{

 \def\Ppi         {\ensuremath{\uppi}\xspace}                 
                  
 \def\Prho        {\ensuremath{\uprho}\xspace}

 \def\Ppsi        {\ensuremath{\uppsi}\xspace}

 \def\PDelta      {\ensuremath{\Delta}\xspace}                 
 \def\PXi         {\ensuremath{\Xi}\xspace}                 
 \def\PLambda     {\ensuremath{\Lambda}\xspace}                 
 \def\PSigma      {\ensuremath{\Sigma}\xspace}                 
 \def\POmega      {\ensuremath{\Omega}\xspace}                 
 \def\PUpsilon    {\ensuremath{\Upsilon}\xspace}

 \def\PB      {\ensuremath{\mathrm{B}}\xspace}                 
                  
 \def\PD      {\ensuremath{\mathrm{D}}\xspace}

 \def\PJ      {\ensuremath{\mathrm{J}}\xspace}                 
 \def\PK      {\ensuremath{\mathrm{K}}\xspace}

 \def\Pb      {\ensuremath{\mathrm{b}}\xspace}                 
 \def\Pc      {\ensuremath{\mathrm{c}}\xspace}                 
 \def\Pd      {\ensuremath{\mathrm{d}}\xspace}                 
 \def\Pe      {\ensuremath{\mathrm{e}}\xspace}

 \def\Pi      {\ensuremath{\mathrm{i}}\xspace}

 \def\Pp      {\ensuremath{\mathrm{p}}\xspace}                 
 \def\Pq      {\ensuremath{\mathrm{q}}\xspace}                 
                  
 \def\Ps      {\ensuremath{\mathrm{s}}\xspace}                 
                  
 \def\Pu      {\ensuremath{\mathrm{u}}\xspace}

 \def\thebaroffset{0.0em}
}
{

 \def\Ppi         {\ensuremath{\pi}\xspace}                 
                  
 \def\Prho        {\ensuremath{\rho}\xspace}

 \def\Ppsi        {\ensuremath{\psi}\xspace}                 
                  
 \mathchardef\PDelta="7101
 \mathchardef\PXi="7104
 \mathchardef\PLambda="7103
 \mathchardef\PSigma="7106
 \mathchardef\POmega="710A
 \mathchardef\PUpsilon="7107
                  
 \def\PB      {\ensuremath{B}\xspace}                 
                  
 \def\PD      {\ensuremath{D}\xspace}

 \def\PJ      {\ensuremath{J}\xspace}                 
 \def\PK      {\ensuremath{K}\xspace}

 \def\Pb      {\ensuremath{b}\xspace}                 
 \def\Pc      {\ensuremath{c}\xspace}                 
 \def\Pd      {\ensuremath{d}\xspace}                 
 \def\Pe      {\ensuremath{e}\xspace}

 \def\Pi      {\ensuremath{i}\xspace}

 \def\Pp      {\ensuremath{p}\xspace}                 
 \def\Pq      {\ensuremath{q}\xspace}                 
                  
 \def\Ps      {\ensuremath{s}\xspace}                 
                  
 \def\Pu      {\ensuremath{u}\xspace}

 \def\thebaroffset{0.18em}
}
\newcommand{\offsetoverline}[2][\thebaroffset]{\kern #1\overline{\kern -#1 #2}}%

\makeatletter
\ifcase \@ptsize \relax
  \newcommand{\miniscule}{\@setfontsize\miniscule{4}{5}}
\or
  \newcommand{\miniscule}{\@setfontsize\miniscule{5}{6}}
\or
  \newcommand{\miniscule}{\@setfontsize\miniscule{5}{6}}
\fi
\makeatother

\DeclareRobustCommand{\optbar}[1]{\shortstack{{\miniscule (\rule[.5ex]{1.25em}{.18mm})}
  \\ [-.7ex] $#1$}}

\def\en         {{\ensuremath{\Pe^-}}\xspace}   
\def\ep         {{\ensuremath{\Pe^+}}\xspace}

\def\quark     {{\ensuremath{\Pq}}\xspace}

\def\uquark    {{\ensuremath{\Pu}}\xspace}

\def\dquark    {{\ensuremath{\Pd}}\xspace}

\def\squark    {{\ensuremath{\Ps}}\xspace}

\def\cquark    {{\ensuremath{\Pc}}\xspace}

\def\bquark    {{\ensuremath{\Pb}}\xspace}

\def\pion   {{\ensuremath{\Ppi}}\xspace}
\def\piz    {{\ensuremath{\pion^0}}\xspace}
\def\pip    {{\ensuremath{\pion^+}}\xspace}
\def\pim    {{\ensuremath{\pion^-}}\xspace}

\def\rhomeson {{\ensuremath{\Prho}}\xspace}

\def\rhom     {{\ensuremath{\rhomeson^-}}\xspace}

\def\kaon    {{\ensuremath{\PK}}\xspace}

\def\KorKbar {\kern \thebaroffset\optbar{\kern -\thebaroffset \PK}{}\xspace}

\def\Km      {{\ensuremath{\kaon^-}}\xspace}

\def\D       {{\ensuremath{\PD}}\xspace}

\def\DorDbar {\kern \thebaroffset\optbar{\kern -\thebaroffset \PD}\xspace}

\def\Dp      {{\ensuremath{\D^+}}\xspace}
\def\Dm      {{\ensuremath{\D^-}}\xspace}

\def\DpDm    {\ensuremath{\Dp {\kern -0.16em \Dm}}\xspace}

\def\B       {{\ensuremath{\PB}}\xspace}

\def\BorBbar {\kern \thebaroffset\optbar{\kern -\thebaroffset \PB}\xspace}

\def\Bd      {{\ensuremath{\B^0}}\xspace}

\def\BdorBdbar {\kern \thebaroffset\optbar{\kern -\thebaroffset \Bd}\xspace}

\def\Bs      {{\ensuremath{\B^0_\squark}}\xspace}

\def\BsorBsbar {\kern \thebaroffset\optbar{\kern -\thebaroffset \Bs}\xspace}

\def\jpsi     {{\ensuremath{{\PJ\mskip -3mu/\mskip -2mu\Ppsi}}}\xspace}

\def\Y#1S{\ensuremath{\PUpsilon{(#1S)}}\xspace}

\def\proton      {{\ensuremath{\Pp}}\xspace}

\def\LorLbar     {\kern \thebaroffset\optbar{\kern -\thebaroffset \PLambda}\xspace}

\def\Xires       {{\ensuremath{\PXi}}\xspace}

\def\Omegares    {{\ensuremath{\POmega}}\xspace}

\def\Xicz        {{\ensuremath{\Xires^0_\cquark}}\xspace}
\def\Xicp        {{\ensuremath{\Xires^+_\cquark}}\xspace}

\def\Omegac      {{\ensuremath{\Omegares^0_\cquark}}\xspace}

\def\Omegab       {{\ensuremath{\Omegares^-_\bquark}}\xspace}

\newcommand{\decay}[2]{\ensuremath{#1\!\to #2}\xspace} 

\def\to                 {\ensuremath{\rightarrow}\xspace}

\def\AT#1     {\ensuremath{A_{\mathrm{T}}^{#1}}\xspace}           

\def\C#1      {\ensuremath{\mathcal{C}_{#1}}\xspace}                       
\def\Cp#1     {\ensuremath{\mathcal{C}_{#1}^{'}}\xspace}                    
\def\Ceff#1   {\ensuremath{\mathcal{C}_{#1}^{\mathrm{(eff)}}}\xspace}        
\def\Cpeff#1  {\ensuremath{\mathcal{C}_{#1}^{'\mathrm{(eff)}}}\xspace}       
\def\Ope#1    {\ensuremath{\mathcal{O}_{#1}}\xspace}                       
\def\Opep#1   {\ensuremath{\mathcal{O}_{#1}^{'}}\xspace}                    

\newcommand{\aunit}[1]{\ensuremath{\text{\,#1}}}       

\newcommand{\tev}{\aunit{Te\kern -0.1em V}\xspace}
\newcommand{\gev}{\aunit{Ge\kern -0.1em V}\xspace}
\newcommand{\mev}{\aunit{Me\kern -0.1em V}\xspace}
\newcommand{\kev}{\aunit{ke\kern -0.1em V}\xspace}
\newcommand{\ev}{\aunit{e\kern -0.1em V}\xspace}
 
\newcommand{\mevc}{\ensuremath{\aunit{Me\kern -0.1em V\!/}c}\xspace}
\newcommand{\gevc}{\ensuremath{\aunit{Ge\kern -0.1em V\!/}c}\xspace}
\newcommand{\mevcc}{\ensuremath{\aunit{Me\kern -0.1em V\!/}c^2}\xspace}
\newcommand{\gevcc}{\ensuremath{\aunit{Ge\kern -0.1em V\!/}c^2}\xspace}

\def\fb   {\ensuremath{\aunit{fb}}\xspace}
\def\invfb   {\ensuremath{\fb^{-1}}\xspace}

\def\ps   {\ensuremath{\aunit{ps}}\xspace}

\newcommand{\chisq}{\ensuremath{\chi^2}\xspace}

\newcommand{\chisqip}{\ensuremath{\chi^2_{\text{IP}}}\xspace}

\def\gsim{{~\raise.15em\hbox{$>$}\kern-.85em
          \lower.35em\hbox{$\sim$}~}\xspace}
\def\lsim{{~\raise.15em\hbox{$<$}\kern-.85em
          \lower.35em\hbox{$\sim$}~}\xspace}

\def\pt         {\ensuremath{p_{\mathrm{T}}}\xspace}

\def\evtgen     {\mbox{\textsc{EvtGen}}\xspace}

\def\geant      {\mbox{\textsc{Geant4}}\xspace}

\def\pythia     {\mbox{\textsc{Pythia}}\xspace}

\def\tell1  {TELL1\xspace}
\def\ukl1   {UKL1\xspace}

\usepackage{cite} 
\usepackage{mciteplus}

\usepackage{longtable} 
\newcommand{\OmegacXX}{\ensuremath{\POmega_\cquark^{**0}}\xspace}
\newcommand{\OmegacXXa}{\ensuremath{\POmega_{\cquark}(3000)^{0}}\xspace}
\newcommand{\OmegacXXb}{\ensuremath{\POmega_{\cquark}(3050)^{0}}\xspace}
\newcommand{\OmegacXXc}{\ensuremath{\POmega_{\cquark}(3065)^{0}}\xspace}
\newcommand{\OmegacXXd}{\ensuremath{\POmega_{\cquark}(3090)^{0}}\xspace}
\newcommand{\OmegacXXe}{\ensuremath{\POmega_{\cquark}(3120)^{0}}\xspace}
\newcommand{\pcal}[1]{\ensuremath{\mathcal{P}_{#1}}\xspace}
\newcommand{\pcaloverXicKpi}[1]{\pcal{#1}}
\newcommand{\pcalXiKpioverOcpi}{\ensuremath{\mathcal{R}}\xspace}

\usepackage[detect-all]{siunitx}
\usepackage{pgfplots}
\pgfplotsset{compat=newest}
\usepgfplotslibrary{groupplots}
\usepgfplotslibrary{polar}
\usepgfplotslibrary{smithchart}
\usepgfplotslibrary{statistics}
\usepgfplotslibrary{dateplot}
\usetikzlibrary{arrows.meta}
\usetikzlibrary{backgrounds}
\usepgfplotslibrary{patchplots}
\usepgfplotslibrary{fillbetween}
\pgfplotsset{%
layers/standard/.define layer set={%
    background,axis background,axis grid,axis ticks,axis lines,axis tick labels,pre main,main,axis descriptions,axis foreground%
}{grid style= {/pgfplots/on layer=axis grid},%
    tick style= {/pgfplots/on layer=axis ticks},%
    axis line style= {/pgfplots/on layer=axis lines},%
    label style= {/pgfplots/on layer=axis descriptions},%
    legend style= {/pgfplots/on layer=axis descriptions},%
    title style= {/pgfplots/on layer=axis descriptions},%
    colorbar style= {/pgfplots/on layer=axis descriptions},%
    ticklabel style= {/pgfplots/on layer=axis tick labels},%
    axis background@ style={/pgfplots/on layer=axis background},%
    3d box foreground style={/pgfplots/on layer=axis foreground},%
    },
}

\graphicspath{
    {figs/}
}

\begin{document}

\renewcommand{\thefootnote}{\fnsymbol{footnote}}
\setcounter{footnote}{1}

\onecolumn

\begin{titlepage}
\pagenumbering{roman}

\vspace*{-1.5cm}
\centerline{\large EUROPEAN ORGANIZATION FOR NUCLEAR RESEARCH (CERN)}
\vspace*{1.5cm}
\noindent
\begin{tabular*}{\linewidth}{lc@{\extracolsep{\fill}}r@{\extracolsep{0pt}}}
\ifthenelse{\boolean{pdflatex}}
{\vspace*{-1.5cm}\mbox{\!\!\!\includegraphics[width=.14\textwidth]{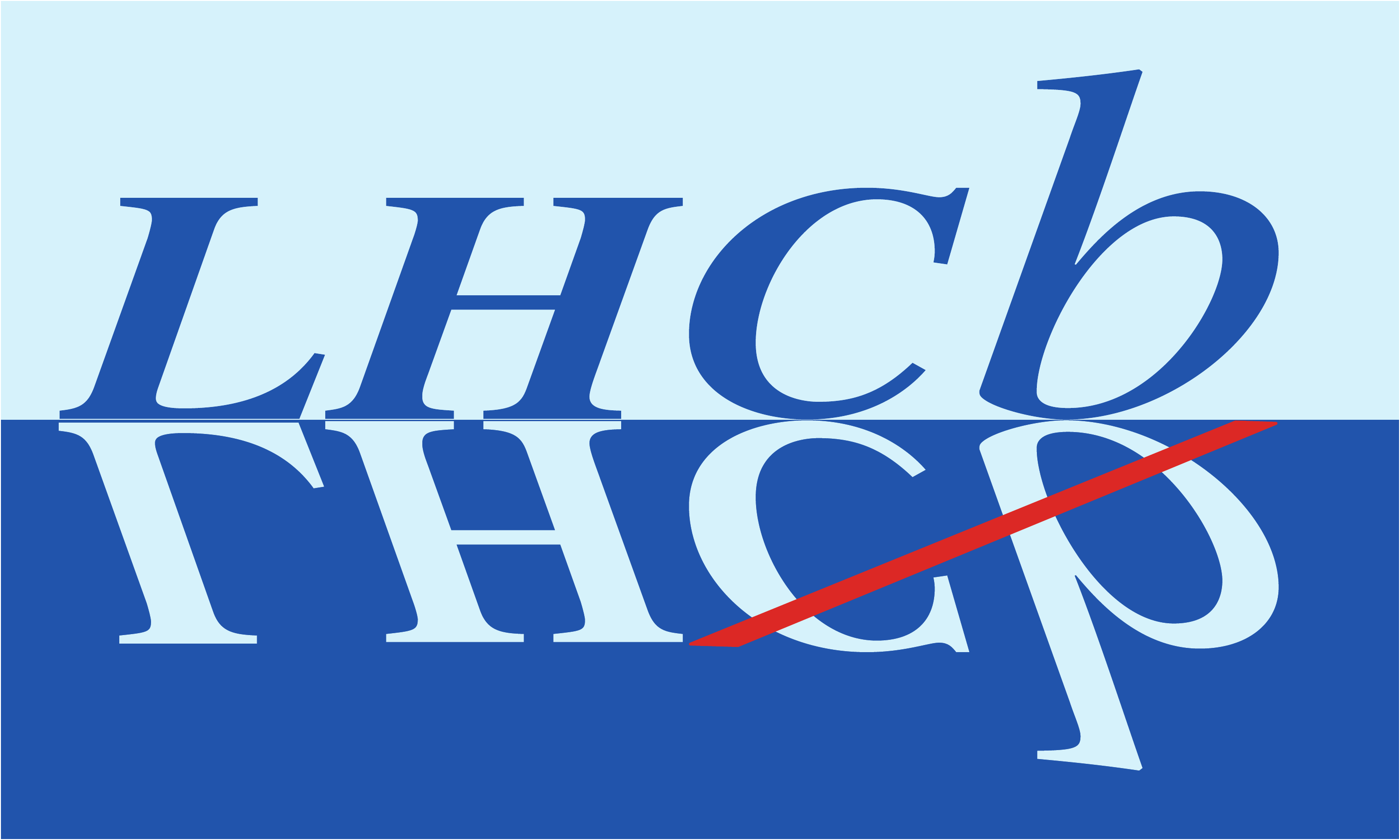}} & &}%
{\vspace*{-1.2cm}\mbox{\!\!\!\includegraphics[width=.12\textwidth]{figs/lhcb-logo.eps}} & &}%
\\
 & & CERN-EP-2021-109 \\  
 & & LHCb-PAPER-2021-012 \\  
 & & 21 September 2021 \\
 & & \\
\end{tabular*}

\vspace*{4.0cm}

{\normalfont\bfseries\boldmath\huge
\begin{center}
  \papertitle 
\end{center}
}

\vspace*{2.0cm}

\begin{center}
\paperauthors\footnote{Authors are listed at the end of this Letter.}
\end{center}

\vspace{\fill}

\begin{abstract}
  \noindent
  The first observation of the \decay{\Omegab}{\Xicp \Km \pim} decay is reported using \proton\proton collision data at centre-of-mass energies of 7, 8 and 13\tev collected by the \lhcb experiment, corresponding to an integrated luminosity of 9\invfb. Four excited \Omegac baryons are observed in the \Xicp \Km mass projection of the \decay{\Omegab}{\Xicp \Km \pim} decays with the significance of each exceeding five standard deviations.
  They coincide with the states previously observed in prompt \proton\proton and $e^+e^-$ production.
  Relative production rates, masses and natural widths of the states are measured, and a test of spin hypotheses is performed. Moreover, the branching ratio of \decay{\Omegab}{\Xicp \Km \pim} is measured relative to the \decay{\Omegab}{\Omegac \pim} decay mode and a precise measurement of the \Omegab mass of \mbox{$6044.3 \pm 1.2 \pm 1.1^{\,+0.19}_{\,-0.22}\mev$} is obtained. 
\end{abstract}

\vspace*{2.0cm}

\begin{center}
  Published in Phys. Rev. D 104 (2021) L091102
\end{center}

\vspace{\fill}

{\footnotesize 
\centerline{\copyright~\papercopyright. \href{\paperlicenceurl}{\paperlicence}.}}
\vspace*{2mm}

\end{titlepage}


\newpage
\setcounter{page}{2}
\mbox{~}

\renewcommand{\thefootnote}{\arabic{footnote}}
\setcounter{footnote}{0}

\cleardoublepage


\pagestyle{plain} 
\setcounter{page}{1}
\pagenumbering{arabic}


\section{Introduction}
\label{sec:Introduction}
The spectrum of the baryons with a single heavy quark $Q\quark\quark^\prime$ ($Q = b$ or $c$ and $\quark,\quark^\prime=\uquark,\dquark$ or \squark) is well classified using the heavy quark-diquark degrees of freedom.
Heavy quark effective theory~\cite{Isgur:1989vq, Isgur:1990yhj, Grinstein:1990mj, Georgi:1990um, Eichten:1989zv, Falk:1990yz, HQET1, HQET2} provides the basis for factoring out the heavy-quark dynamics up to corrections of the first order of $1/m_Q$, where $m_Q$ is the heavy quark mass.
Therefore, the observation of new baryons and measurements of their properties provide information about the role played by diquarks in 
baryons, and can also help to tune tetraquark and pentaquark models.

In  recent years, the \lhcb experiment has made numerous contributions to the spectroscopy of heavy baryons by observing several new states~\cite{LHCb-PAPER-2020-004, LHCb-PAPER-2020-032, LHCb-PAPER-2019-045, LHCb-PAPER-2019-042, LHCb-PAPER-2019-025, LHCb-PAPER-2018-032, LHCb-PAPER-2018-013, LHCb-PAPER-2017-002}. Among them, the spectrum of excited \Omegac baryons has drawn special attention.
Five new excited narrow \Omegac states, 
herein denoted \OmegacXX,
and promptly produced in proton-proton~(\proton\proton) collisions, have been observed in the \Xicp\Km mass spectrum~\cite{LHCb-PAPER-2017-002,Yelton:2017qxg}.

Many theoretical approaches including
potential models,
QCD sum rules, and lattice QCD 
predict the $\OmegacXX$ spectrum and 
interpret the newly discovered states as orbitally or
radially excited \Omegac states~\cite{
  Chiladze:1997ev,Wang:2017vnc, 
  Padmanath:2017lng,
  Cheng:2017ove,
  Capstick:1986bm,Huang:2017dwn, Zhao:2017fov,Chen:2017gnu,Luo:2021dvj,Galkin:2020iat,Roberts:2007ni,Shah:2016mig,Yoshida:2015tia,Karliner:2017kfm,Wang:2017hej,
  Agaev:2017lip,Chen:2017sci,Chen:2015kpa,Wang:2017zjw},
while a few studies suggest that some of them may be either molecular states or pentaquarks\cite{Chen:2017xat, Kim:2017jpx, An:2017lwg, Ali:2017wsf, Montana:2017kjw, Debastiani:2017ewu, Santopinto:2018ljf}.
Most of the predictions propose the mass ordering of the states, 
while widths and relative production rates remain unexploited on the theoretical side.
Seven excited $P$-wave \Omegac baryons are expected: five $\lambda$-mode excited states where the constituent \cquark quark and the $ss$ diquark are in a $P$-wave, 
and two $\rho$-mode excited states where the two \squark quarks are in a $P$-wave.
One of the most popular interpretations is that the observed \OmegacXX states correspond to the five $\lambda$-mode excited \Omegac baryons with quantum numbers $J^P = 1/2^-,  1/2^-,  3/2^-,  3/2^-$, and  $5/2^-$.
The determination of the spin-parity quantum numbers of the \OmegacXX states would help to discriminate between the proposed models and to probe the internal structure of the baryons.

This letter presents the first observation of the \OmegacXX states produced in exclusive \Omegab decays.
These are studied in the previously unobserved \decay{\Omegab}{\Xicp\Km\pim} decays~\cite{Debastiani:2018adr,Chua:2019yqh}, where the \Xicp baryons are reconstructed in the \proton\Km\pip final state.
The mass of the $\Omegab$ baryon has been measured in decays to the $\Omegac\pim$ and $\POmega^- \jpsi$ final states.
The new decay mode \decay{\Omegab}{\Xicp\Km\pim} is a prominent reaction to measure also the \Omegab mass due to a multi-particle final state and smaller phase space with respect to the $\Omegac\pim$ mode.\footnote{Unless otherwise stated, charge-conjugate processes are implicitly included, and natural units with $\hbar = c = 1$ are used throughout.}
The analysis is based on samples of \proton\proton collision data at centre-of-mass energies of
$\sqrt s = 7, 8$ and 13\tev, corresponding to an integrated luminosity of 9\invfb.

\section{Detector and simulation}
\label{sec:Detector}

The \lhcb detector~\cite{LHCb-DP-2008-001,LHCb-DP-2014-002} is a single-arm forward
spectrometer covering the \mbox{pseudorapidity} range $2<\eta <5$,
designed for the study of particles containing \bquark or \cquark
quarks. The detector includes a high-precision tracking system
consisting of a silicon-strip vertex detector surrounding the \proton\proton
interaction region, a large-area silicon-strip detector located
upstream of a dipole magnet with a bending power of about
$4\,\mathrm{Tm}$, and three stations of silicon-strip detectors together with straw drift tubes placed downstream of the magnet.
Simulation is necessary to train a multivariate algorithm used to suppress background, model shapes of mass distributions, and calculate efficiencies.
In the simulation, \proton\proton collisions are generated using \pythia~\cite{Sjostrand:2007gs,*Sjostrand:2006za}
with a specific \lhcb configuration~\cite{LHCb-PROC-2010-056}.
Decays of unstable particles are described by \evtgen~\cite{Lange:2001uf}.
The interaction of the generated particles with the detector is implemented using the \geant toolkit~\cite{Allison:2006ve, *Agostinelli:2002hh} as described in Ref~\cite{LHCb-PROC-2011-006}.

\section{Selection of $\boldmath\textbf\Omegab \to \Xicp\Km\pim$ decays}
\label{sec:Selec}

The \Xicp candidates are formed by combining three tracks that are detached from any primary \proton\proton interaction vertex (PV) in the event. A good-quality vertex fit is required to select tracks originating from the same secondary vertex. The \Omegab candidates are selected by combining the \Xicp candidate with two tracks identified as a \Km and a \pim meson.
Loose particle identification (PID) requirements are applied to all five final-state tracks in order to reduce background. The \Omegab candidates are required to have a transverse momentum $\pt > 3.5\gev$ and are constrained to originate from the PV by requiring a small \chisqip, where \chisqip is defined as the difference in the vertex-fit \chisq of a given PV reconstructed with and without the candidate under consideration.
The \Omegab decay time is required to be larger than 0.2\ps, making the overlap with the prompt sample analysed in Ref.~\cite{LHCb-PAPER-2017-002} negligible.

A boosted decision tree (BDT) classifier, implemented using the TMVA toolkit~\cite{Hocker:2007ht,*TMVA4}, is used to further reduce the background.
Variables found to provide good discrimination between signal and background are: the PID information and \pt of the final-state tracks, the \Xicp \pt, the \Xicp and \Omegab \chisqip, the \Xicp and \Omegab vertex-fit \chisq, the \Omegab flight-distance significance, defined as the measured flight distance divided by its uncertainty, and the cosine of the \Xicp and \Omegab direction angles. The direction angle is defined as the angle between the \Xicp (\Omegab) momentum and the vector joining the PV and the \Xicp (\Omegab) decay vertex.
The training of the BDT classifier is performed using simulated samples as signal and data as background
separately for Run~1 and Run~2 data samples.
The candidates used for the background sample are in the \numrange{6200}{6300}\mev range of the \Xicp\Km\pim mass spectrum, which is not populated by partially reconstructed \Omegab decays. 
The optimal selection criterion on the BDT response is found by maximising the figure of merit $\epsilon/(5/2 + \sqrt{B_P})$~\cite{Punzi:2003bu}, where $\epsilon$ is the signal efficiency in simulation, and $B_P$ is the number of \Xicp\Km\pim candidates in the mass region $6200 < m(\Xicp\Km\pim) < 6256\mev$, 
roughly matching the expected number of background events in the \Omegab mass window.
Roughly $4\%$ of selected events contain more than one candidate and are removed.
Finally, a kinematic fit~\cite{Hulsbergen:2005pu} is applied to the \Omegab decays to improve the mass resolution where the \Xicp candidate mass is constrained to its known value~\cite{PDG2020}, and the \Omegab candidate is constrained to originate from its associated PV, defined as the PV to which the impact parameter of the combination of two-track and $\Xicp$ candidate is the smallest.

\begin{figure}
    \centering
    \includegraphics[width=0.48\textwidth]{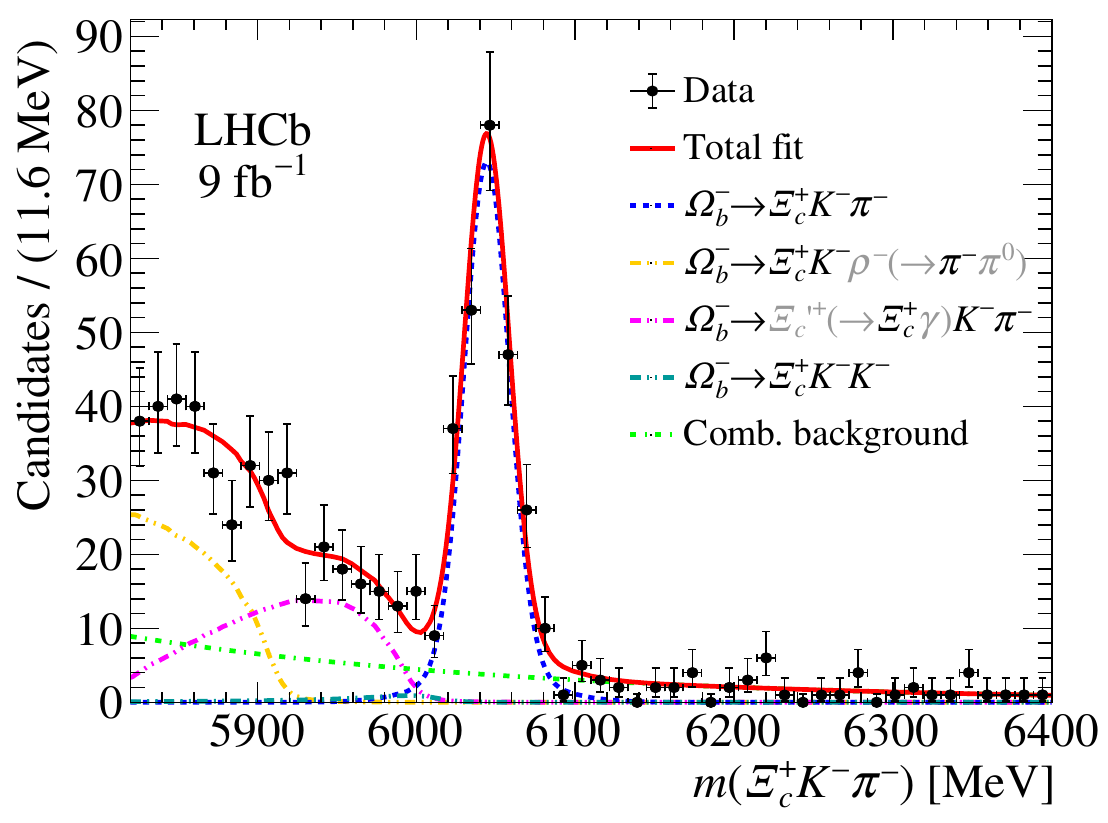}
    \includegraphics[width=0.48\textwidth]{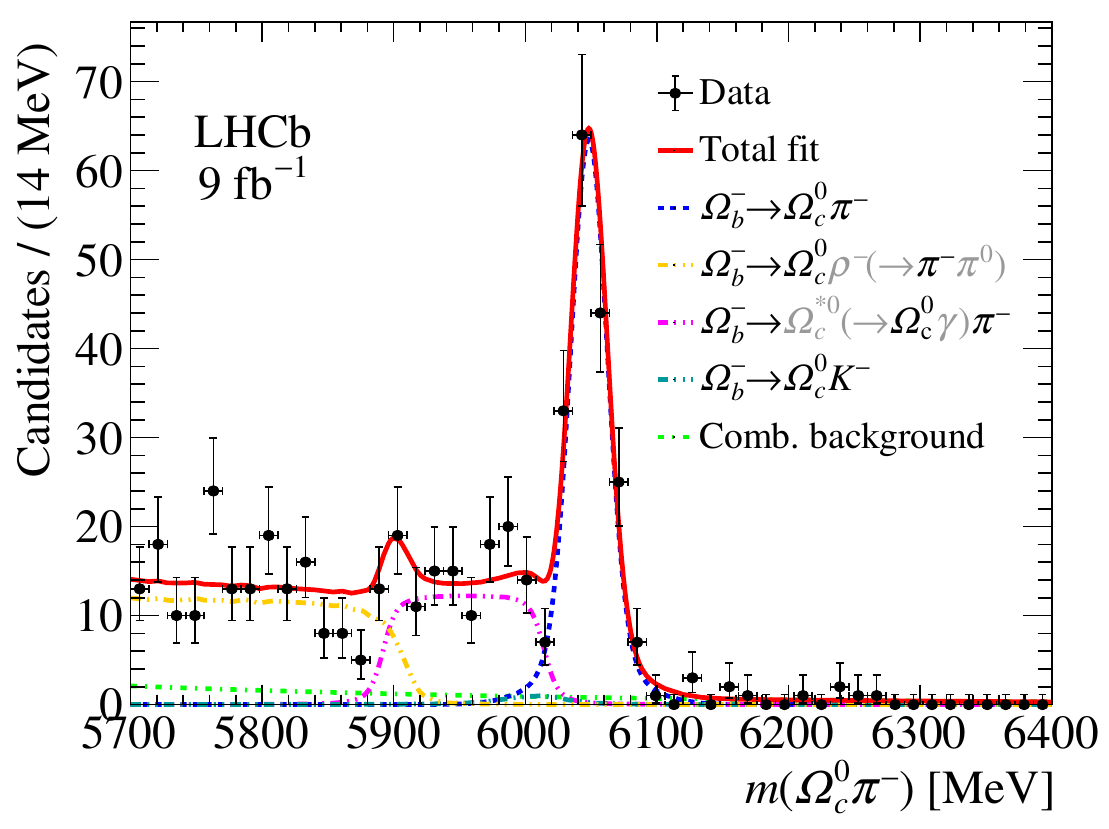}
    \caption{Distribution of the reconstructed invariant mass (left) $m(\Xicp\Km\pim)$ with $\Xicp \to \proton\Km\pip$ and (right) $m(\Omegac\pim)$ with $\Omegac \to \proton\Km\Km\pip$ for all candidates passing the selection requirements. The black symbols show the data. The result of a fit is overlaid (solid red line).
    The missing particles in partially reconstructed decays are indicated in grey in the legends.}
    \label{fig:ObFit}
\end{figure}

\begin{figure}
  \centering
  \includegraphics[width=0.6\linewidth]{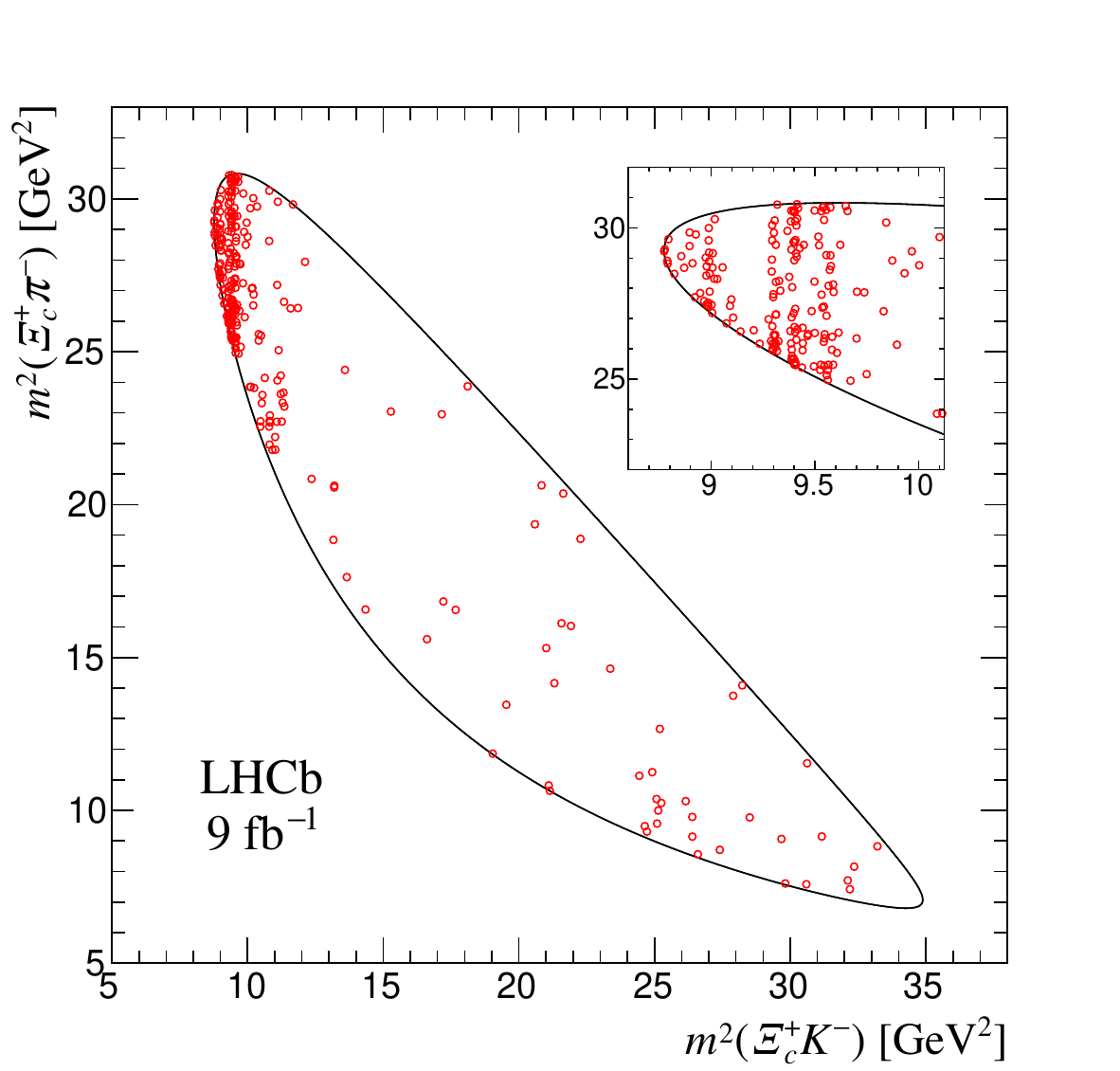}
  \caption{Dalitz plot distribution of \decay{\Omegab}{\Xicp\Km\pim} candidates in the signal region, including background contributions. The inset shows an expanded view of the upper left corner where the vertical bands correspond to excited \Omegac states.}
  \label{fig:Dalitz}
\end{figure}

The resulting \Xicp\Km\pim mass spectrum is shown in Fig.~\ref{fig:ObFit}~(left) and an extended unbinned maximum-likelihood fit is performed.
The signal shape is modelled by the combination of two Gaussian functions with a common mean, where the ratios of the resolutions and yields between the functions are fixed according to the simulation. The main sources of background are due to the partially reconstructed decays \decay{\Omegab}{\Xicp\Km\rhom (\to \pim \piz)} and \decay{\Omegab}{\PXi_{\cquark}^{\prime+}(\to\Xicp\gamma)\Km\pim}, where the \piz and $\gamma$ are not reconstructed.
The combinatorial background shape is fixed according to a 
wrong-sign sample, consisting of \Xicp\Km\pip combinations
processed in the same way as the right-sign \Xicp\Km\pim combinations. 
The shape of the partially reconstructed decays is taken from simulated samples generated using the RapidSim package~\cite{Cowan:2016tnm}. The shape of misidentified decays \decay{\Omegab}{\Xicp\Km\Km} is fixed based on simulation. The yield ratio $\text{N}_{\Xicp\Km\Km}/\text{N}_{\Xicp\Km\pim}$ is fixed to 2.8\% based on $|V_{us}|^2/|V_{ud}|^2 \approx 5\%$ corrected by the difference in reconstruction efficiency and the phase space.
The fit returns a combined mass resolution of $17.9 \pm 1.3 \mev$, a yield of $\text{N}_{\Xicp\Km\pim} = 240\pm 17$ and an \Omegab mass, $m(\Omegab) = 6044.3 \pm 1.2\;\mev$, where the uncertainty is statistical only (see Table~\ref{tab:final_results}).
The Dalitz plot distribution of the candidates, with a mass within two standard deviations of the \Omegab peak, is shown in Fig.~\ref{fig:Dalitz}. Excited \Omegac baryons appear in the \Xicp\Km projection while no excited \Xicz states are clearly visible in the \Xicp\pim system.

The branching fraction of \decay{\Omegab}{\Xicp\Km\pim} decays is measured relative to the normalisation channel \decay{\Omegab}{\Omegac\pim}, with \decay{\Omegac}{\proton\Km\Km\pip}.
Similar selection requirements as the \decay{\Omegab}{\Xicp\Km\pim} mode are applied to the \decay{\Omegab}{\Omegac\pim} candidates. The selections of the two decay modes differ in the requirements applied to the invariant mass of the \proton\Km\pip and \proton\Km\Km\pip systems to select \Xicp and \Omegac candidates, respectively.
A kinematic fit is applied to the \Omegab decay where the \Omegac candidate mass is constrained to its known value~\cite{PDG2020}.
The two largest background components are due to the partially reconstructed decays \decay{\Omegab}{\Omegac\rhom (\to \pim\piz) }, and \decay{\Omegab}{\POmega_\cquark^{*0}(\to \Omegac\gamma)\pim}.
The result of an unbinned maximum-likelihood fit is overlaid to the data in Fig.~\ref{fig:ObFit}~(right). 
All decays are modelled in the same way as for the \decay{\Omegab}{\Xicp\Km\pim} channel.
The combinatorial background shape is fixed according to the projection of the \Omegac sidebands in the \Omegac\pim mass spectrum, where the \Omegac sidebands are defined as the \numrange{2650}{2670} and \numrange{2720}{2740}\mev ranges in the \proton\Km\Km\pip invariant mass distribution.
The yield of reconstructed \Omegab candidates is $\text{N}_{\Omegac\pim} = 174 \pm 14$, and the mass resolution is $18.4 \pm 1.5\mev$.

The ratio of branching fractions is obtained as
\begin{align} \nonumber
  \pcalXiKpioverOcpi &\equiv 
\frac{\mathcal{B}(\decay{\Omegab}{\Xicp\Km\pim})\,\mathcal{B}(\decay{\Xicp}{\proton\Km\pip})}{\mathcal{B}(\Omegab \to \Omegac\pim)\,\mathcal{B}(\decay{\Omegac}{\proton\Km\Km\pip})}  =  1.35 \pm 0.11 \ ,
\end{align}
\noindent which is calculated from the ratio of efficiency-corrected yields, where the error is statistical only (see Table~\ref{tab:final_results}).

\section{The $\boldmath\textbf{\Xicp\Km}$ mass spectrum}
\label{sec:XicKSpec}

\begin{figure}
   \centering
    \includegraphics[width=0.6\textwidth]{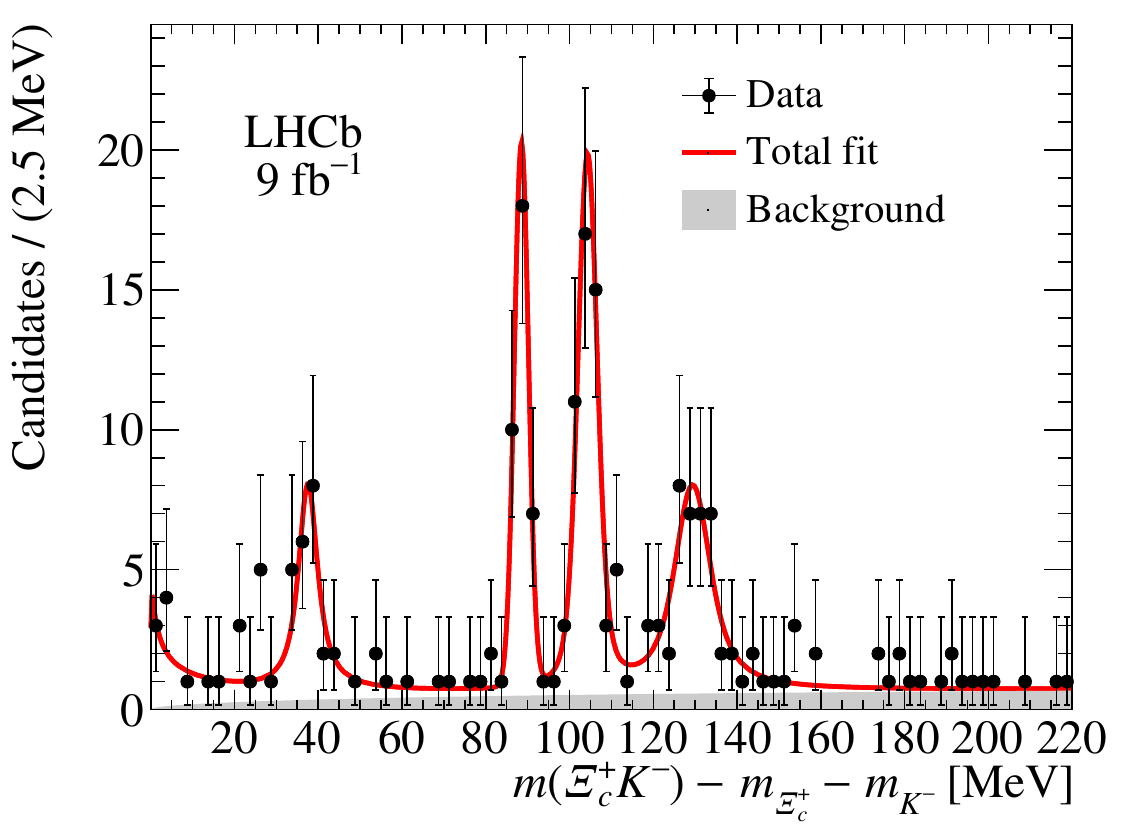}
    \caption{Distribution of the reconstructed mass difference between the \Xicp\Km invariant mass and the \Xicp and \Km masses.
    The four peaking structures are consistent with being the previously observed \OmegacXXa, \OmegacXXb, \OmegacXXc and \OmegacXXd baryons. The distribution shows an enhancement at the threshold, as seen in the previous analysis~\cite{LHCb-PAPER-2017-002}.
    The total fit is overlaid in red. The background distribution (grey shaded area) is the combination of the combinatorial and nonresonant \Xicp\Km backgrounds. 
    }
    \label{fig:XicKmass}
\end{figure}

A search for excited \Omegac baryons is performed in the \Xicp\Km mass projection
of \decay{\Omegab}{\Xicp\Km\pim} candidates.
In order to increase the selection efficiency of the \OmegacXX states, 
an additional BDT classifier is deployed for the study of the \Xicp\Km spectrum, where a sample of simulated \decay{\Omegab}{ \Xicp\Km \pim} decays, with an additional requirement of \mbox{$m(\Xicp\Km) < 3.3\gev$}, is used as the signal sample. For the background, the upper region of the \Xicp\Km\pim mass distribution is used, as in the previous BDT classifier.
After the optimization of the BDT response, the \Omegab candidates with a mass within two standard deviations of the \Omegab peak are selected. Figure~\ref{fig:XicKmass} shows the distribution of the mass difference $\Delta M \equiv m(\Xicp\Km) - m_\Xicp -m_\Km$, where $m(\Xicp\Km)$ is the invariant mass of the \Xicp\Km system, and $m_\Xicp$ and $m_\Km$ are the world averages of the \Xicp and \Km masses, respectively~\cite{PDG2020}. Four narrow peaking structures are clearly visible close to the \Xicp \Km kinematic threshold.

An extended maximum-likelihood fit is performed
to the $\Delta M$ distribution, where each signal is modelled by an $S$-wave relativistic Breit-Wigner (BW) function multiplied by the phase space function and convolved with a Gaussian function to describe the mass resolution.
The widths and masses of the relativistic BW functions vary freely. The background consists of two components:
the combinatorial background under the \Omegab signal peak (Fig.~\ref{fig:ObFit}~(left)) and the nonresonant \Xicp\Km component. The former (combinatorial) is modelled by projecting the \Omegab sideband into the \Xicp\Km invariant mass distribution and the latter (nonresonant \Xicp\Km) according to phase space. While the shapes of the two contributions and the yield of the combinatorial component are fixed, the yield of the nonresonant background can vary freely.
The $\Xicp\Km$ spectrum also features an excess at the \Xicp\Km mass threshold which is modelled by an $S$-wave BW component.
Fit results superimposed to the data are shown in Fig.~\ref{fig:XicKmass}.
The yields attributed to the four peaks are $24 \pm 7$, $33 \pm 6$, $51 \pm 8$, and $41 \pm 9$ respectively.
The resulting BW parameters of the four signals, which are listed in Table~\ref{tab:final_results},
are consistent with those of the previously observed \OmegacXXa, \OmegacXXb, \OmegacXXc and \OmegacXXd baryons~\cite{LHCb-PAPER-2017-002}.
The natural width of the \OmegacXXb is consistent with zero, therefore an upper limit is set.
In order to determine the significance of the peaking structures, another fit is performed by fixing the masses and widths of the \OmegacXX states to the previously measured values~\cite{LHCb-PAPER-2017-002}. Therefore, the statistical significance of each peak is calculated using $\sqrt{2 \Delta(\text{NLL})}$, where $\Delta(\text{NLL})$ is the variation of the fit log-likelihood when the corresponding BW function is excluded from the reference fit model. The local significance exceeds six standard deviations $(6\,\sigma)$ for each of the four main states. 
For the threshold structure, the null hypothesis of the background fluctuation is tested using the likelihood ratio of two fits. The $p$-value expressed in standard deviations using the one-sided convention corresponds to $4.3\,\sigma$ after systematic uncertainties are accounted for.
Finally, the production rate of the \OmegacXX states relative to the \decay{\Omegab}{\Xicp\Km\pim} mode is defined as
\begin{align}
    \pcaloverXicKpi{\OmegacXX} \equiv \frac{\mathcal{B}(\Omegab \to \OmegacXX\pim)\,\mathcal{B}(\OmegacXX \to \Xicp \Km)}{\mathcal{B}(\Omegab \to \Xicp\Km\pim)}\,.
\end{align}
The rate is measured for the \OmegacXXa, \OmegacXXb, \OmegacXXc and \OmegacXXd, and an upper limit on the production of the \OmegacXXe state is set.
The results are reported in Table~\ref{tab:final_results} with the statistical error computed using the binomial distribution.

\section{Spin test for the excited $\boldmath\textbf{\Omegac}$ baryons}
\label{sec:AngAna}

\begin{figure}
    \centering
    \includegraphics[width=0.48\textwidth]{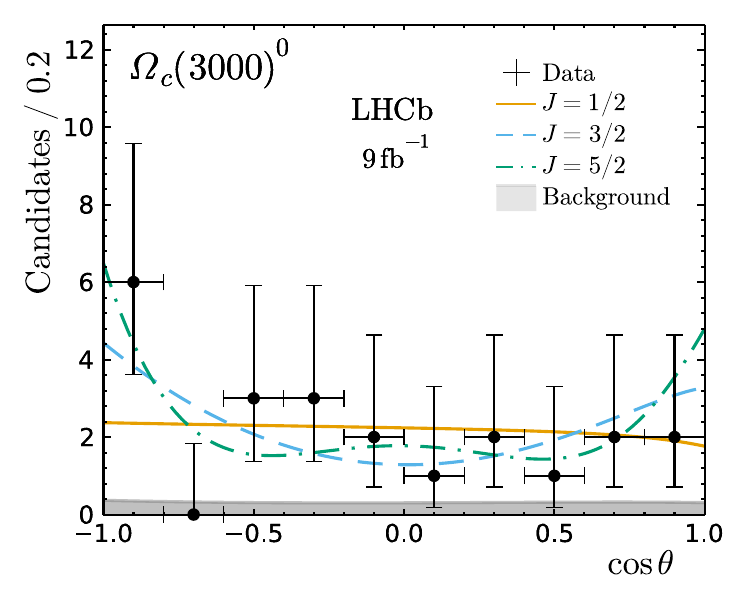}\,
    \includegraphics[width=0.48\textwidth]{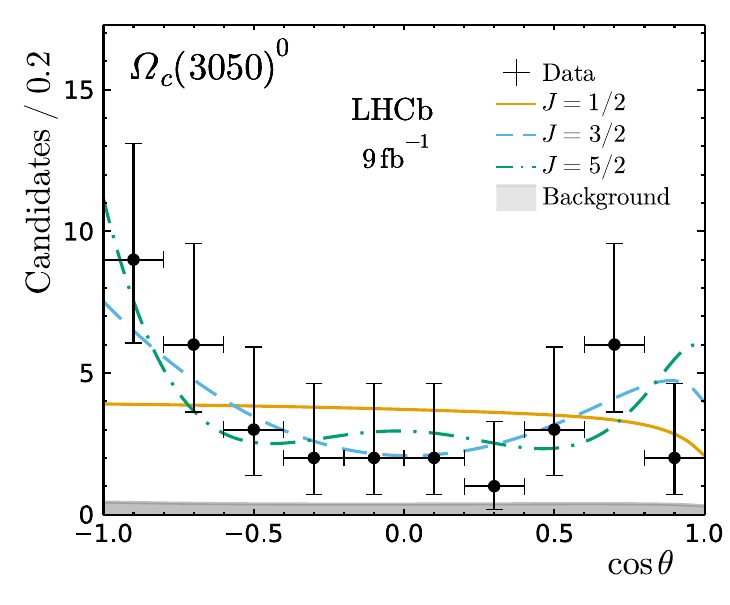}\,
    \includegraphics[width=0.48\textwidth]{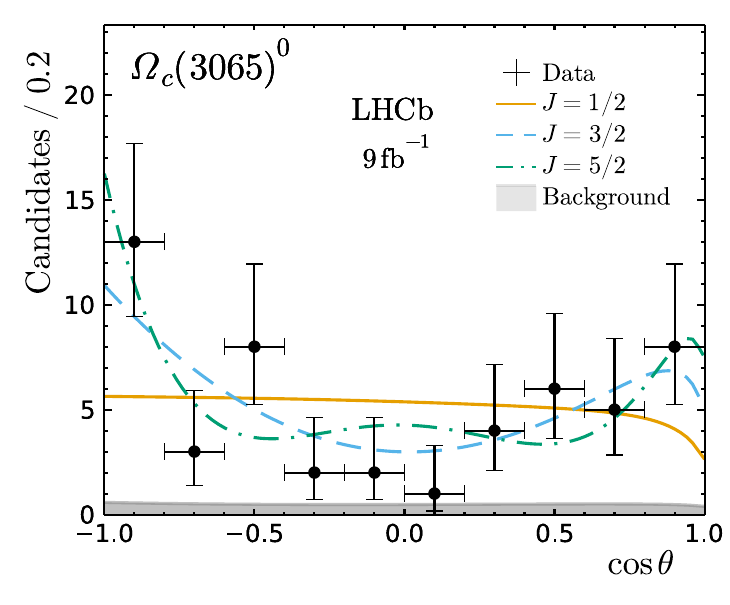}\,
    \includegraphics[width=0.48\textwidth]{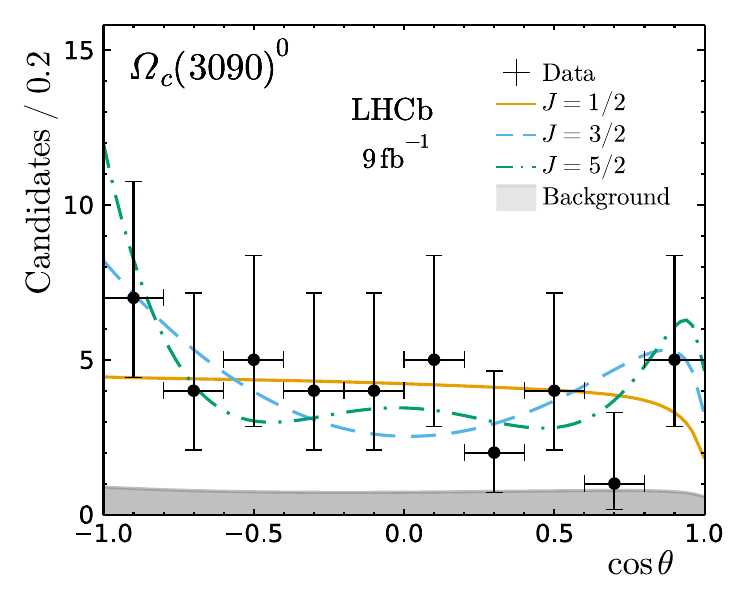}
    \caption{Distributions of the \OmegacXX helicity angle ($\theta$) in the \Omegab decay.
    Solid, dashed and dot-dashed lines indicate the expectations under the spin hypotheses, $J=1/2$, $3/2$, and $5/2$, respectively. The grey shaded area shows the cumulative distribution of the combinatorial and nonresonant \Xicp\Km backgrounds.}
    \label{fig:angular.distribution}
\end{figure}

In order to probe the spin of the \OmegacXX baryons, the distribution
of the helicity angle in the $\decay{\Omegab}{\OmegacXX(\rightarrow \Xicp\Km)\pim}$ decay is studied.
The helicity angle $\theta$
is defined as the angle between the $\vec{p}_\Km$ and the $-\vec{p}_\pim$ directions in the $\Xicp\Km$ rest frame, where $\vec{p}$ is the momentum of the meson.
The spin projection of the $\OmegacXX$ baryon in the direction of the $\pim$ meson is limited to $1/2$ as it is produced in the $\decay{\Omegab}{\OmegacXX \pim}$ decay.
Additionally, it cannot exceed $1/2$ in the direction of either decay product, $\Xicp$ or $\Km$, due to their spins. Therefore, the angular distribution for a $\OmegacXX$ state with spin $J$ is given as
\begin{align} \label{eq:Ij}
    I_{J}(\cos\theta) = \frac{(2J+1)}{2} \left(\left|d^J_{1/2,-1/2}(\cos\theta)\right|^2 + \left|d^J_{1/2,+1/2}(\cos\theta)\right|^2\right),
\end{align}
where $d_{\nu,\lambda}^J$ is the Wigner $d$-function. The first (second) index, $\nu$ ($\lambda$), gives the spin projections of the $\OmegacXX$ in the direction opposite to the pion (kaon) momentum, $-\vec{p}_{\pim}$ ($-\vec{p}_{\Km}$), in the $\Xicp\Km$ rest frame.
The angular distributions are not affected by a possible polarization of the $\Omegab$ baryon since its production angles are integrated over. 
The \OmegacXX candidates are selected in the small nonoverlapping regions around the peaks.
The $\cos{\theta}$ distributions for the \OmegacXX states are shown in Fig.~\ref{fig:angular.distribution}.
The $\OmegacXXb$ and $\OmegacXXc$ distributions show an enhancement at $\cos\theta=-1$,
hinting at a preference for a spin larger than $J=1/2$.

The expectations for the angular density function, $D_J(\cos\theta)$, shown by the colored lines in Fig.~\ref{fig:angular.distribution},
are calculated as a sum of the signal PDF and the two background components (combinatorial and nonresonant \Xicp\Km) by
\begin{align}
    D_J(\cos\theta) \equiv
      f_s I_J(\cos\theta) \epsilon(\cos\theta) +
      f_{b} B_1(\cos\theta)  + (1-f_s-f_{b}) B_2(\cos\theta) \epsilon(\cos\theta)\,,
\end{align}
where $f_s$ and $f_{b}$ are the fractions of the signal and the combinatorial background fixed according to the result of the mass fit.
The angular distribution for the combinatorial background, $B_1(\cos\theta)$, is fixed by selecting candidates in the $\Xicp\Km\pim$ mass range above the $\Omegab$ peak.
A flat distribution is assumed for nonresonant background, $B_2(\cos\theta)$. The efficiency, $\epsilon(\cos\theta)$, is calculated separately for each signal region using simulation. The efficiency maps are combined according to the fraction of the signal candidates in the corresponding data-taking periods. The efficiency for the helicity angle is calculated by convolving the efficiency map with the \OmegacXX line-shape profile.  The fall of the curves at $\cos\theta=1$ indicates the smaller reconstruction efficiency for candidates with a low momentum $\Km$ in the $\Omegab$ rest frame.
Discrimination of different spin hypotheses is based on the likelihood-ratio test statistic,
\begin{align} \label{eq:test.statistics}
    t_{H_J|H_{J'}} &= \frac{1}{N}\sum_{i=1}^{N} \log \left[ D_{H_J}(\cos\theta_i) / D_{H_{J'}}(\cos\theta_i) \right],
\end{align}
where $H_{J}$ and $H_{J'}$ are the compared hypotheses for the state to have spin $J$ and $J'$, respectively, $N$ is the number of candidates in the mass region around the peak.
The test statistic $\vec t\,^{(\text{data})} = (t^{(\text{data})}_{J=1/2|J=3/2}, t^{(\text{data})}_{J=3/2|J=5/2})$ is evaluated in data and compared to the $t$ distribution
in simulated pseudoexperiments. A set of $20\,000$ pseudoexperiments with the number of signal and background events obtained from data are simulated for each spin hypothesis and for every $\OmegacXX$ state.
The two-dimensional distribution of $t$ is well described by the multivariate normal distribution
from which we extract the covariance matrix and the two-dimensional mean, $t^{(\text{mean})}$.
The $p$-value in the double-tailed convention is calculated by $\exp(-r^2/2)$, where $r$ is the Mahalanobis distance~\cite{DeMaesschalck:2000xyz} between $\vec t\,^{(\text{data})}$ and $\vec t\,^{(\text{mean})}$.
All results are summarized in Table~\ref{tab:final_results}.
The significance of the rejection of the $J=1/2$ hypothesis for $\OmegacXXb$ and $\OmegacXXc$ is $2.2\,\sigma$ and $3.6\,\sigma$ respectively, including systematic effects listed in the next section.
The combined hypothesis of the four peaks to have quantum numbers in the order $1/2$, $1/2$, $3/2$, $3/2$ is tested and rejected with a significance of $3.5\,\sigma$. 

\section{Systematic uncertainties}
\label{sec:Syst}

Various systematic uncertainties for each observable are considered, where the largest deviation from the default model on every source is used.
A summary of the  systematic uncertainties is provided in the supplemental material~\cite{LHCb-PAPER-2021-012-supp}. The uncertainties from different sources are combined in quadrature. A source of systematic uncertainty is determined from varying components of the \Omegab fit model. The helicity couplings of the partially reconstructed decays in the \Omegac\pim invariant mass spectrum are modified as well as the shape used to describe the signal peaks.
The uncertainty in the yield of misidentified decays is quantified by varying the fractional contribution by $\pm40\%$ relative to the default value. In simulation, the \Xicp \to \proton \Km \pip Dalitz plot is generated according to phase space and a binned weighting is performed to match the data. A systematic uncertainty is found by varying the binning scheme.

The uncertainty in the mass measurements due to momentum calibration is determined following Ref.~\cite{LHCb-PAPER-2013-011} as $\pm0.03\%$ of the energy released in the decay.
The PID variables in simulation are corrected in order to match the PID performance in data. To calculate an uncertainty, a modified weighting is applied to the PID variables. 
For the uncertainty in the \Omegab kinematics, the \pt and $\eta$ of the \Omegab candidates, as well as the track multiplicity in simulation, are weighted according to data.
Several alternative models are considered for \Xicp\Km fit.
Firstly, the resolution of each Gaussian function is varied by $\pm10\%$.
In addition, different orbital angular momenta ($L = 1,2$) are tested along with the variation of the Blatt-Weisskopf factors~\cite{Blatt:1952ije,VonHippel:1972fg} from $1.5$ to $5\gev^{-1}$.
A constant-width BW approximation and the scattering-length approximation are probed for the threshold structure. Lastly, for each signal peak, interference with neighbours and the nonresonant \Xicp\Km background is tested. The full list of results including systematic uncertainties are listed in Table~\ref{tab:final_results}. 

\section{Summary and conclusion}\label{sec:Conc}
In summary, data collected by the LHCb experiment at centre-of-mass energies 7, 8 and 13\tev corresponding to an integrated luminosity of 9\invfb are used to observe the new decay mode \decay{\Omegab}{\Xicp\Km\pim} and to measure its branching fraction relative to the \decay{\Omegab}{\Omegac\pim} decay mode. A precise measurement of the \Omegab mass, $m(\Omegab) = 6044.3 \pm 1.2 \pm 1.1^{\,+0.19}_{\,-0.22}\mev$, is obtained
where the first uncertainty is statistical, the second is systematic, and the third asymmetric error is due to the uncertainty in the $\Xicp$ mass. Averaging with the previous \lhcb measurements~\cite{LHCb-PAPER-2016-008,LHCb-PAPER-2012-048}, taking correlated systematic uncertainties into account, gives a mass of $m(\Omegab) = 6044.8 \pm 1.3 \mev$, which is the most precise to date.

The investigation of the \Xicp\Km mass spectrum has revealed four excited \Omegac baryons, \OmegacXXa, \OmegacXXb, \OmegacXXc, \OmegacXXd, and a threshold enhancement as also seen in Ref.~\cite{LHCb-PAPER-2017-002}.
The \OmegacXXe state is not observed, therefore an upper limit on its production rate is set by scanning the likelihood profile, $\pcaloverXicKpi{\OmegacXXe} < 0.03$ at $95\%$ confidence level~(CL).
Measurements of the \OmegacXX masses and widths, 
together with an upper limit of $\Gamma_{\OmegacXXb}<1.6\mev$ at 95\% CL
are reported. Their spin assignments are tested based on the distribution of the helicity angle in the decay chain $\Omegab \to \OmegacXX \pim, \; \OmegacXX\to\Xicp\Km$.
Significance values of excluding the $J=1/2$ spin hypothesis for $\OmegacXXb$ and \OmegacXXc are $2.2\,\sigma$ and $3.6\,\sigma$, respectively. All results are summarised in Table~\ref{tab:final_results}.
The combined hypothesis on the spin of the four peaks in the order $J = 1/2$, $1/2$, $3/2$, $3/2$,
as proposed in several works~\cite{Karliner:2017kfm, Padmanath:2017lng, Wang:2017zjw}, is rejected with a $p$-value corresponding to $3.5$ standard deviations once systematic uncertainties are taken into account.

The results of the angular analysis together to the absence of the \OmegacXXe state in the \Xicp\Km spectrum in \Omegab decays and in \ep\en collisions at Belle~\cite{Yelton:2017qxg}, suggest that the interpretation of the five peaks observed in Ref.~\cite{LHCb-PAPER-2017-002} as $\lambda$-mode excited states might be invalid.
In such a scenario, only the four peaks observed in this analysis would be $\lambda$-mode excitations (with quantum numbers $J=1/2$, $3/2$, $3/2$, and $5/2$) and a spin $1/2$ $\lambda$-mode state would be still to be observed. 
The nonobservation of the \OmegacXXe baryon
would be consistent with the state being either one of the $2S$ doublet, decaying to $\Xicp\Km$ in $P$-wave~\cite{Karliner:2017kfm, Galkin:2020iat}, or a $\rho$-mode $P$-wave excitation with spin $3/2^-$ that requires $D$-wave between \Xicp and \Km.
Finally, the $\Xicp\Km$ spectrum also features an excess at the \Xicp\Km mass threshold. An analogous enhancement was observed in the inclusive $\Xicp\Km$ spectrum~\cite{LHCb-PAPER-2017-002} and 
interpreted as the partially reconstructed decay $\OmegacXXc\to \PXi_\cquark^{\prime +} (\to \Xicp\gamma) \Km$
with the photon escaping detection. However, such an explanation does not hold here, given that the partially reconstructed decay $\Omegab \to \PXi_\cquark^{\prime +} \Km \pim$
does not populate the mass region selected for the exclusive \decay{\Omegab}{\Xicp\Km\pim} decay.
While the current data do not provide enough sensitivity to
determine the parameters of the structure, such as the mass, natural
width and spin, future data acquired with the upgraded LHCb detector
will provide insights to establish its nature.

\begin{table}[]
    \centering
    \caption{Results on the \Omegab mass, relative branching fraction of the $\Xicp\Km\pim$ decay mode, measured mass differences ($\Delta M$), masses ($m$), natural widths ($\Gamma$) and production fraction~($\cal{P}$) of \OmegacXX baryons where the first uncertainty is statistical and the second systematic. The third asymmetric uncertainty on the \Omegab and \OmegacXX masses is due to the uncertainty in the \Xicp mass. Upper limits are given for the width of the \OmegacXXb state and the production rate of the \OmegacXXe baryon, which are measured to be consistent with zero. The results of the spin analysis are also listed ($J$ rejection).}
    \label{tab:final_results}
    \begin{tabular}{lc | c }
    State & Observable & Measurement\\
    \hline
    \multirow{2}{*}{\Omegab}
            &                 $m$ &  $6044.3 \pm 1.2 \pm 1.1^{\,+0.19}_{\,-0.22}\mev$ \\
            &  $\pcalXiKpioverOcpi$ &                      $1.35 \pm 0.11 \pm 0.05$ \\
    \hline\hline
    Threshold
            &        \multirow{2}{*}{Significance} &   \multirow{2}{*}{$4.3\,\sigma$} \\
    structure & & \\
    \hline
    \multirow{6}{*}{\OmegacXXa}
            &        Significance &                                   $6.2\,\sigma$ \\
            &          $\Delta M$ &                       $37.6\pm0.9\pm 0.9\mev$ \\
            &                 $m$ & $2999.2 \pm 0.9 \pm 0.9 ^{\,+0.19}_{\,-0.22}\mev$ \\
            &            $\Gamma$ &                      $4.8\pm 2.1\pm 2.5 \mev$ \\
            &           $\pcal{}$ &                     $0.11 \pm 0.02 \pm 0.04 $ \\
            &       $J$ rejection &             $0.5\,\sigma \,(J=1/2), 0.8\,\sigma \,(J=3/2), 0.4\,\sigma \,(J=5/2)$\\
    \hline
    \multirow{6}{*}{\OmegacXXb}
            &        Significance &                                   $9.9\,\sigma$ \\
            &          $\Delta M$ &                      $88.5\pm0.3\pm 0.2 \mev$ \\
            &                $m $ & $3050.1 \pm 0.3 \pm 0.2 ^{\,+0.19}_{\,-0.22}\mev$ \\
            &            $\Gamma$ &                          $< 1.6\mev$, 95\% CL \\
            &           $\pcal{}$ &                      $0.15 \pm 0.02 \pm 0.02$ \\
            &       $J$ rejection &             
            $2.2\,\sigma\, (J=1/2), 0.1\,\sigma\, (J=3/2), 1.2\,\sigma \,(J=5/2)$\\
    \hline
    \multirow{6}{*}{\OmegacXXc}
            &        Significance &                                  $11.9\,\sigma$ \\
            &          $\Delta M$ &                     $104.3\pm0.4\pm 0.4 \mev$ \\
            &                $m $ & $3065.9 \pm 0.4 \pm 0.4 ^{\,+0.19}_{\,-0.22}\mev$ \\
            & $\Gamma$            &                    $1.7 \pm 1.0 \pm 0.5 \mev$ \\
            &           $\pcal{}$ &                      $0.23 \pm 0.02 \pm 0.02$ \\
            &       $J$ rejection &             
            $3.6\,\sigma\, (J=1/2), 0.6\,\sigma\, (J=3/2), 1.2\,\sigma \,(J=5/2)$\\
    \hline
    \multirow{6}{*}{\OmegacXXd}
            &        Significance &                                   $7.8\,\sigma$ \\
            &          $\Delta M$ &                     $129.4\pm1.1\pm 1.0 \mev$ \\
            &                $m $ & $3091.0 \pm 1.1 \pm 1.0 ^{\,+0.19}_{\,-0.22}\mev$ \\
            &            $\Gamma$ &                    $7.4 \pm 3.1 \pm 2.8 \mev$ \\
            &           $\pcal{}$ &                      $0.19 \pm 0.02 \pm 0.04$ \\
            &       $J$ rejection &             
            $0.3\,\sigma\, (J=1/2), 0.8\,\sigma\, (J=3/2), 0.5\,\sigma \,(J=5/2)$\\
    \hline
    \OmegacXXe
            &           $\pcal{}$ &                      $<0.03$, $95\%$ CL \\
    \hline   
        \end{tabular}
\end{table}

\clearpage

\section*{Acknowledgements}
%
%
\noindent We express our gratitude to our colleagues in the CERN
accelerator departments for the excellent performance of the LHC. We
thank the technical and administrative staff at the LHCb
institutes.
We acknowledge support from CERN and from the national agencies:
CAPES, CNPq, FAPERJ and FINEP (Brazil); 
MOST and NSFC (China); 
CNRS/IN2P3 (France); 
BMBF, DFG and MPG (Germany); 
INFN (Italy); 
NWO (Netherlands); 
MNiSW and NCN (Poland); 
MEN/IFA (Romania); 
MSHE (Russia); 
MICINN (Spain); 
SNSF and SER (Switzerland); 
NASU (Ukraine); 
STFC (United Kingdom); 
DOE NP and NSF (USA).
We acknowledge the computing resources that are provided by CERN, IN2P3
(France), KIT and DESY (Germany), INFN (Italy), SURF (Netherlands),
PIC (Spain), GridPP (United Kingdom), RRCKI and Yandex
LLC (Russia), CSCS (Switzerland), IFIN-HH (Romania), CBPF (Brazil),
PL-GRID (Poland) and OSC (USA).
We are indebted to the communities behind the multiple open-source
software packages on which we depend.
Individual groups or members have received support from
AvH Foundation (Germany);
EPLANET, Marie Sk\l{}odowska-Curie Actions and ERC (European Union);
A*MIDEX, ANR, Labex P2IO and OCEVU, and R\'{e}gion Auvergne-Rh\^{o}ne-Alpes (France);
Key Research Program of Frontier Sciences of CAS, CAS PIFI, CAS CCEPP, 
Fundamental Research Funds for the Central Universities, 
and Sci. \& Tech. Program of Guangzhou (China);
RFBR, RSF and Yandex LLC (Russia);
GVA, XuntaGal and GENCAT (Spain);
the Royal Society
and the Leverhulme Trust (United Kingdom).

\addcontentsline{toc}{section}{References}
\bibliographystyle{LHCb}
\bibliography{main,standard,LHCb-PAPER,LHCb-CONF,LHCb-DP,LHCb-TDR}

\ifx\mcitethebibliography\mciteundefinedmacro
\PackageError{LHCb.bst}{mciteplus.sty has not been loaded}
{This bibstyle requires the use of the mciteplus package.}\fi
\providecommand{\href}[2]{#2}
\begin{mcitethebibliography}{10}
\mciteSetBstSublistMode{n}
\mciteSetBstMaxWidthForm{subitem}{\alph{mcitesubitemcount})}
\mciteSetBstSublistLabelBeginEnd{\mcitemaxwidthsubitemform\space}
{\relax}{\relax}

\bibitem{Isgur:1989vq}
N.~Isgur and M.~B. Wise, \ifthenelse{\boolean{articletitles}}{\emph{{Weak
  decays of heavy mesons in the static quark approximation}},
  }{}\href{https://doi.org/10.1016/0370-2693(89)90566-2}{Phys.\ Lett.\
  \textbf{B232} (1989) 113}\relax
\mciteBstWouldAddEndPuncttrue
\mciteSetBstMidEndSepPunct{\mcitedefaultmidpunct}
{\mcitedefaultendpunct}{\mcitedefaultseppunct}\relax
\EndOfBibitem
\bibitem{Isgur:1990yhj}
N.~Isgur and M.~B. Wise, \ifthenelse{\boolean{articletitles}}{\emph{{Weak
  transition form factors between heavy mesons}},
  }{}\href{https://doi.org/10.1016/0370-2693(90)91219-2}{Phys.\ Lett.\
  \textbf{B237} (1990) 527}\relax
\mciteBstWouldAddEndPuncttrue
\mciteSetBstMidEndSepPunct{\mcitedefaultmidpunct}
{\mcitedefaultendpunct}{\mcitedefaultseppunct}\relax
\EndOfBibitem
\bibitem{Grinstein:1990mj}
B.~Grinstein, \ifthenelse{\boolean{articletitles}}{\emph{{The static quark
  effective theory}},
  }{}\href{https://doi.org/10.1016/0550-3213(90)90349-I}{Nucl.\ Phys.\
  \textbf{B339} (1990) 253}\relax
\mciteBstWouldAddEndPuncttrue
\mciteSetBstMidEndSepPunct{\mcitedefaultmidpunct}
{\mcitedefaultendpunct}{\mcitedefaultseppunct}\relax
\EndOfBibitem
\bibitem{Georgi:1990um}
H.~Georgi, \ifthenelse{\boolean{articletitles}}{\emph{{An effective field
  theory for heavy quarks at low energies}},
  }{}\href{https://doi.org/10.1016/0370-2693(90)91128-X}{Phys.\ Lett.\
  \textbf{B240} (1990) 447}\relax
\mciteBstWouldAddEndPuncttrue
\mciteSetBstMidEndSepPunct{\mcitedefaultmidpunct}
{\mcitedefaultendpunct}{\mcitedefaultseppunct}\relax
\EndOfBibitem
\bibitem{Eichten:1989zv}
E.~Eichten and B.~R. Hill, \ifthenelse{\boolean{articletitles}}{\emph{{An
  effective field theory for the calculation of matrix elements involving heavy
  quarks}}, }{}\href{https://doi.org/10.1016/0370-2693(90)92049-O}{Phys.\
  Lett.\  \textbf{B234} (1990) 511}\relax
\mciteBstWouldAddEndPuncttrue
\mciteSetBstMidEndSepPunct{\mcitedefaultmidpunct}
{\mcitedefaultendpunct}{\mcitedefaultseppunct}\relax
\EndOfBibitem
\bibitem{Falk:1990yz}
A.~F. Falk, H.~Georgi, B.~Grinstein, and M.~B. Wise,
  \ifthenelse{\boolean{articletitles}}{\emph{{Heavy meson form factors from
  QCD}}, }{}\href{https://doi.org/10.1016/0550-3213(90)90591-Z}{Nucl.\ Phys.\
  \textbf{B343} (1990) 1}\relax
\mciteBstWouldAddEndPuncttrue
\mciteSetBstMidEndSepPunct{\mcitedefaultmidpunct}
{\mcitedefaultendpunct}{\mcitedefaultseppunct}\relax
\EndOfBibitem
\bibitem{HQET1}
A.~G. Grozin, \ifthenelse{\boolean{articletitles}}{\emph{Introduction to the
  heavy quark effective theory},
  }{}\href{http://arxiv.org/abs/hep-ph/9908366}{{\normalfont\ttfamily
  arXiv:hep-ph/9908366}}\relax
\mciteBstWouldAddEndPuncttrue
\mciteSetBstMidEndSepPunct{\mcitedefaultmidpunct}
{\mcitedefaultendpunct}{\mcitedefaultseppunct}\relax
\EndOfBibitem
\bibitem{HQET2}
T.~Mannel, \ifthenelse{\boolean{articletitles}}{\emph{{Effective theory for
  heavy quarks}}, }{}\href{https://doi.org/10.1007/BFb0104296}{Lect.\ Notes
  Phys.\  \textbf{479} (1997) 387},
  \href{http://arxiv.org/abs/hep-ph/9606299}{{\normalfont\ttfamily
  arXiv:hep-ph/9606299}}\relax
\mciteBstWouldAddEndPuncttrue
\mciteSetBstMidEndSepPunct{\mcitedefaultmidpunct}
{\mcitedefaultendpunct}{\mcitedefaultseppunct}\relax
\EndOfBibitem
\bibitem{LHCb-PAPER-2020-004}
LHCb collaboration, R.~Aaij {\em et~al.},
  \ifthenelse{\boolean{articletitles}}{\emph{{Observation of new \Xicz baryons
  decaying to $\Lc\Km$}},
  }{}\href{https://doi.org/10.1103/PhysRevLett.124.222001}{Phys.\ Rev.\ Lett.\
  \textbf{124} (2020) 222001},
  \href{http://arxiv.org/abs/2003.13649}{{\normalfont\ttfamily
  arXiv:2003.13649}}\relax
\mciteBstWouldAddEndPuncttrue
\mciteSetBstMidEndSepPunct{\mcitedefaultmidpunct}
{\mcitedefaultendpunct}{\mcitedefaultseppunct}\relax
\EndOfBibitem
\bibitem{LHCb-PAPER-2020-032}
LHCb collaboration, R.~Aaij {\em et~al.},
  \ifthenelse{\boolean{articletitles}}{\emph{{Observation of a new $\Xibz$
  state}}, }{}\href{https://doi.org/10.1103/PhysRevD.103.012004}{Phys.\ Rev.\
  \textbf{D103} (2021) 012004},
  \href{http://arxiv.org/abs/2010.14485}{{\normalfont\ttfamily
  arXiv:2010.14485}}\relax
\mciteBstWouldAddEndPuncttrue
\mciteSetBstMidEndSepPunct{\mcitedefaultmidpunct}
{\mcitedefaultendpunct}{\mcitedefaultseppunct}\relax
\EndOfBibitem
\bibitem{LHCb-PAPER-2019-045}
LHCb collaboration, R.~Aaij {\em et~al.},
  \ifthenelse{\boolean{articletitles}}{\emph{{Observation of a new baryon state
  in the $\Lb\pip\pim$ mass spectrum}},
  }{}\href{https://doi.org/10.1007/JHEP06(2020)136}{JHEP \textbf{06} (2020)
  136}, \href{http://arxiv.org/abs/2002.05112}{{\normalfont\ttfamily
  arXiv:2002.05112}}\relax
\mciteBstWouldAddEndPuncttrue
\mciteSetBstMidEndSepPunct{\mcitedefaultmidpunct}
{\mcitedefaultendpunct}{\mcitedefaultseppunct}\relax
\EndOfBibitem
\bibitem{LHCb-PAPER-2019-042}
LHCb collaboration, R.~Aaij {\em et~al.},
  \ifthenelse{\boolean{articletitles}}{\emph{{First observation of excited
  $\Omegares_b^-$ states}},
  }{}\href{https://doi.org/10.1103/PhysRevLett.124.082002}{Phys.\ Rev.\ Lett.\
  \textbf{124} (2020) 082002},
  \href{http://arxiv.org/abs/2001.00851}{{\normalfont\ttfamily
  arXiv:2001.00851}}\relax
\mciteBstWouldAddEndPuncttrue
\mciteSetBstMidEndSepPunct{\mcitedefaultmidpunct}
{\mcitedefaultendpunct}{\mcitedefaultseppunct}\relax
\EndOfBibitem
\bibitem{LHCb-PAPER-2019-025}
LHCb collaboration, R.~Aaij {\em et~al.},
  \ifthenelse{\boolean{articletitles}}{\emph{{Observation of new resonances in
  the \mbox{\Lb\pip\pim} system}},
  }{}\href{https://doi.org/10.1103/PhysRevLett.123.152001}{Phys.\ Rev.\ Lett.\
  \textbf{123} (2019) 152001},
  \href{http://arxiv.org/abs/1907.13598}{{\normalfont\ttfamily
  arXiv:1907.13598}}\relax
\mciteBstWouldAddEndPuncttrue
\mciteSetBstMidEndSepPunct{\mcitedefaultmidpunct}
{\mcitedefaultendpunct}{\mcitedefaultseppunct}\relax
\EndOfBibitem
\bibitem{LHCb-PAPER-2018-032}
LHCb collaboration, R.~Aaij {\em et~al.},
  \ifthenelse{\boolean{articletitles}}{\emph{{Observation of two resonances in
  the $\Lb\pipm$ systems and precise measurement of $\Sigmares_b^\pm$ and
  $\Sigmares_b^{\ast\pm}$ properties}},
  }{}\href{https://doi.org/10.1103/PhysRevLett.122.012001}{Phys.\ Rev.\ Lett.\
  \textbf{122} (2019) 012001},
  \href{http://arxiv.org/abs/1809.07752}{{\normalfont\ttfamily
  arXiv:1809.07752}}\relax
\mciteBstWouldAddEndPuncttrue
\mciteSetBstMidEndSepPunct{\mcitedefaultmidpunct}
{\mcitedefaultendpunct}{\mcitedefaultseppunct}\relax
\EndOfBibitem
\bibitem{LHCb-PAPER-2018-013}
LHCb collaboration, R.~Aaij {\em et~al.},
  \ifthenelse{\boolean{articletitles}}{\emph{{Observation of a new \Xibm
  resonance}}, }{}\href{https://doi.org/10.1103/PhysRevLett.121.072002}{Phys.\
  Rev.\ Lett.\  \textbf{121} (2018) 072002},
  \href{http://arxiv.org/abs/1805.09418}{{\normalfont\ttfamily
  arXiv:1805.09418}}\relax
\mciteBstWouldAddEndPuncttrue
\mciteSetBstMidEndSepPunct{\mcitedefaultmidpunct}
{\mcitedefaultendpunct}{\mcitedefaultseppunct}\relax
\EndOfBibitem
\bibitem{LHCb-PAPER-2017-002}
LHCb collaboration, R.~Aaij {\em et~al.},
  \ifthenelse{\boolean{articletitles}}{\emph{{Observation of five new narrow
  $\Omegac$ states decaying to $\Xicp\Km$}},
  }{}\href{https://doi.org/10.1103/PhysRevLett.118.182001}{Phys.\ Rev.\ Lett.\
  \textbf{118} (2017) 182001},
  \href{http://arxiv.org/abs/1703.04639}{{\normalfont\ttfamily
  arXiv:1703.04639}}\relax
\mciteBstWouldAddEndPuncttrue
\mciteSetBstMidEndSepPunct{\mcitedefaultmidpunct}
{\mcitedefaultendpunct}{\mcitedefaultseppunct}\relax
\EndOfBibitem
\bibitem{Yelton:2017qxg}
Belle collaboration, J.~Yelton {\em et~al.},
  \ifthenelse{\boolean{articletitles}}{\emph{{Observation of excited
  $\POmega_\cquark$ charmed baryons in $e^+e^-$ collisions}},
  }{}\href{https://doi.org/10.1103/PhysRevD.97.051102}{Phys.\ Rev.\
  \textbf{D97} (2018) 051102},
  \href{http://arxiv.org/abs/1711.07927}{{\normalfont\ttfamily
  arXiv:1711.07927}}\relax
\mciteBstWouldAddEndPuncttrue
\mciteSetBstMidEndSepPunct{\mcitedefaultmidpunct}
{\mcitedefaultendpunct}{\mcitedefaultseppunct}\relax
\EndOfBibitem
\bibitem{Chiladze:1997ev}
G.~Chiladze and A.~F. Falk,
  \ifthenelse{\boolean{articletitles}}{\emph{{Phenomenology of new baryons with
  charm and strangeness}},
  }{}\href{https://doi.org/10.1103/PhysRevD.56.R6738}{Phys.\ Rev.\
  \textbf{D56} (1997) R6738},
  \href{http://arxiv.org/abs/hep-ph/9707507}{{\normalfont\ttfamily
  arXiv:hep-ph/9707507}}\relax
\mciteBstWouldAddEndPuncttrue
\mciteSetBstMidEndSepPunct{\mcitedefaultmidpunct}
{\mcitedefaultendpunct}{\mcitedefaultseppunct}\relax
\EndOfBibitem
\bibitem{Wang:2017vnc}
W.~Wang and R.-L. Zhu,
  \ifthenelse{\boolean{articletitles}}{\emph{{Interpretation of the newly
  observed $\POmega_\cquark^0$ resonances}},
  }{}\href{https://doi.org/10.1103/PhysRevD.96.014024}{Phys.\ Rev.\
  \textbf{D96} (2017) 014024},
  \href{http://arxiv.org/abs/1704.00179}{{\normalfont\ttfamily
  arXiv:1704.00179}}\relax
\mciteBstWouldAddEndPuncttrue
\mciteSetBstMidEndSepPunct{\mcitedefaultmidpunct}
{\mcitedefaultendpunct}{\mcitedefaultseppunct}\relax
\EndOfBibitem
\bibitem{Padmanath:2017lng}
M.~Padmanath and N.~Mathur, \ifthenelse{\boolean{articletitles}}{\emph{{Quantum
  numbers of recently discovered $\POmega^{0}_{\cquark}$ baryons from lattice
  QCD}}, }{}\href{https://doi.org/10.1103/PhysRevLett.119.042001}{Phys.\ Rev.\
  Lett.\  \textbf{119} (2017) 042001},
  \href{http://arxiv.org/abs/1704.00259}{{\normalfont\ttfamily
  arXiv:1704.00259}}\relax
\mciteBstWouldAddEndPuncttrue
\mciteSetBstMidEndSepPunct{\mcitedefaultmidpunct}
{\mcitedefaultendpunct}{\mcitedefaultseppunct}\relax
\EndOfBibitem
\bibitem{Cheng:2017ove}
H.-Y. Cheng and C.-W. Chiang,
  \ifthenelse{\boolean{articletitles}}{\emph{{Quantum numbers of $\POmega_c$
  states and other charmed baryons}},
  }{}\href{https://doi.org/10.1103/PhysRevD.95.094018}{Phys.\ Rev.\
  \textbf{D95} (2017) 094018},
  \href{http://arxiv.org/abs/1704.00396}{{\normalfont\ttfamily
  arXiv:1704.00396}}\relax
\mciteBstWouldAddEndPuncttrue
\mciteSetBstMidEndSepPunct{\mcitedefaultmidpunct}
{\mcitedefaultendpunct}{\mcitedefaultseppunct}\relax
\EndOfBibitem
\bibitem{Capstick:1986bm}
S.~Capstick and N.~Isgur, \ifthenelse{\boolean{articletitles}}{\emph{{Baryons
  in a relativized quark model with chromodynamics}},
  }{}\href{https://doi.org/10.1103/PhysRevD.34.2809}{Phys.\ Rev.\  \textbf{D34}
  (1986) 2809}\relax
\mciteBstWouldAddEndPuncttrue
\mciteSetBstMidEndSepPunct{\mcitedefaultmidpunct}
{\mcitedefaultendpunct}{\mcitedefaultseppunct}\relax
\EndOfBibitem
\bibitem{Huang:2017dwn}
H.~Huang, J.~Ping, and F.~Wang,
  \ifthenelse{\boolean{articletitles}}{\emph{{Investigating the excited
  $\POmega^{0}_{c}$ states through $\PXi_{c}\bar \PK$ and $\PXi^{'}_{c}\bar
  \PK$ decay channels}},
  }{}\href{https://doi.org/10.1103/PhysRevD.97.034027}{Phys.\ Rev.\
  \textbf{D97} (2018) 034027},
  \href{http://arxiv.org/abs/1704.01421}{{\normalfont\ttfamily
  arXiv:1704.01421}}\relax
\mciteBstWouldAddEndPuncttrue
\mciteSetBstMidEndSepPunct{\mcitedefaultmidpunct}
{\mcitedefaultendpunct}{\mcitedefaultseppunct}\relax
\EndOfBibitem
\bibitem{Zhao:2017fov}
Z.~Zhao, D.-D. Ye, and A.~Zhang,
  \ifthenelse{\boolean{articletitles}}{\emph{{Hadronic decay properties of
  newly observed $\POmega_c$ baryons}},
  }{}\href{https://doi.org/10.1103/PhysRevD.95.114024}{Phys.\ Rev.\
  \textbf{D95} (2017) 114024},
  \href{http://arxiv.org/abs/1704.02688}{{\normalfont\ttfamily
  arXiv:1704.02688}}\relax
\mciteBstWouldAddEndPuncttrue
\mciteSetBstMidEndSepPunct{\mcitedefaultmidpunct}
{\mcitedefaultendpunct}{\mcitedefaultseppunct}\relax
\EndOfBibitem
\bibitem{Chen:2017gnu}
B.~Chen and X.~Liu, \ifthenelse{\boolean{articletitles}}{\emph{{New
  $\POmega_c^0$ baryons discovered by LHCb as the members of $1P$ and $2S$
  states}}, }{}\href{https://doi.org/10.1103/PhysRevD.96.094015}{Phys.\ Rev.\
  \textbf{D96} (2017) 094015},
  \href{http://arxiv.org/abs/1704.02583}{{\normalfont\ttfamily
  arXiv:1704.02583}}\relax
\mciteBstWouldAddEndPuncttrue
\mciteSetBstMidEndSepPunct{\mcitedefaultmidpunct}
{\mcitedefaultendpunct}{\mcitedefaultseppunct}\relax
\EndOfBibitem
\bibitem{Luo:2021dvj}
S.-Q. Luo, B.~Chen, X.~Liu, and T.~Matsuki,
  \ifthenelse{\boolean{articletitles}}{\emph{{Predicting a new resonance as
  charmed-strange baryonic analog of $D^*_{s0}$(2317)}},
  }{}\href{https://doi.org/10.1103/PhysRevD.103.074027}{Phys.\ Rev.\
  \textbf{D103} (2021) 074027},
  \href{http://arxiv.org/abs/2102.00679}{{\normalfont\ttfamily
  arXiv:2102.00679}}\relax
\mciteBstWouldAddEndPuncttrue
\mciteSetBstMidEndSepPunct{\mcitedefaultmidpunct}
{\mcitedefaultendpunct}{\mcitedefaultseppunct}\relax
\EndOfBibitem
\bibitem{Galkin:2020iat}
V.~O. Galkin and R.~N. Faustov,
  \ifthenelse{\boolean{articletitles}}{\emph{{Heavy baryon spectroscopy}},
  }{}\href{https://doi.org/10.1134/S1063779620040292}{Phys.\ Part.\ Nucl.\
  \textbf{51} (2020) 661}\relax
\mciteBstWouldAddEndPuncttrue
\mciteSetBstMidEndSepPunct{\mcitedefaultmidpunct}
{\mcitedefaultendpunct}{\mcitedefaultseppunct}\relax
\EndOfBibitem
\bibitem{Roberts:2007ni}
W.~Roberts and M.~Pervin, \ifthenelse{\boolean{articletitles}}{\emph{{Heavy
  baryons in a quark model}},
  }{}\href{https://doi.org/10.1142/S0217751X08041219}{Int.\ J.\ Mod.\ Phys.\
  \textbf{A23} (2008) 2817},
  \href{http://arxiv.org/abs/0711.2492}{{\normalfont\ttfamily
  arXiv:0711.2492}}\relax
\mciteBstWouldAddEndPuncttrue
\mciteSetBstMidEndSepPunct{\mcitedefaultmidpunct}
{\mcitedefaultendpunct}{\mcitedefaultseppunct}\relax
\EndOfBibitem
\bibitem{Shah:2016mig}
Z.~Shah, K.~Thakkar, A.~Kumar~Rai, and P.~C. Vinodkumar,
  \ifthenelse{\boolean{articletitles}}{\emph{{Excited state mass spectra of
  singly charmed baryons}},
  }{}\href{https://doi.org/10.1140/epja/i2016-16313-9}{Eur.\ Phys.\ J.\
  \textbf{A52} (2016) 313},
  \href{http://arxiv.org/abs/1602.06384}{{\normalfont\ttfamily
  arXiv:1602.06384}}\relax
\mciteBstWouldAddEndPuncttrue
\mciteSetBstMidEndSepPunct{\mcitedefaultmidpunct}
{\mcitedefaultendpunct}{\mcitedefaultseppunct}\relax
\EndOfBibitem
\bibitem{Yoshida:2015tia}
T.~Yoshida {\em et~al.}, \ifthenelse{\boolean{articletitles}}{\emph{{Spectrum
  of heavy baryons in the quark model}},
  }{}\href{https://doi.org/10.1103/PhysRevD.92.114029}{Phys.\ Rev.\
  \textbf{D92} (2015) 114029},
  \href{http://arxiv.org/abs/1510.01067}{{\normalfont\ttfamily
  arXiv:1510.01067}}\relax
\mciteBstWouldAddEndPuncttrue
\mciteSetBstMidEndSepPunct{\mcitedefaultmidpunct}
{\mcitedefaultendpunct}{\mcitedefaultseppunct}\relax
\EndOfBibitem
\bibitem{Karliner:2017kfm}
M.~Karliner and J.~L. Rosner, \ifthenelse{\boolean{articletitles}}{\emph{{Very
  narrow excited $\POmega_\cquark$ baryons}},
  }{}\href{https://doi.org/10.1103/PhysRevD.95.114012}{Phys.\ Rev.\
  \textbf{D95} (2017) 114012},
  \href{http://arxiv.org/abs/1703.07774}{{\normalfont\ttfamily
  arXiv:1703.07774}}\relax
\mciteBstWouldAddEndPuncttrue
\mciteSetBstMidEndSepPunct{\mcitedefaultmidpunct}
{\mcitedefaultendpunct}{\mcitedefaultseppunct}\relax
\EndOfBibitem
\bibitem{Wang:2017hej}
K.-L. Wang, L.-Y. Xiao, X.-H. Zhong, and Q.~Zhao,
  \ifthenelse{\boolean{articletitles}}{\emph{{Understanding the newly observed
  $\POmega_c$ states through their decays}},
  }{}\href{https://doi.org/10.1103/PhysRevD.95.116010}{Phys.\ Rev.\
  \textbf{D95} (2017) 116010},
  \href{http://arxiv.org/abs/1703.09130}{{\normalfont\ttfamily
  arXiv:1703.09130}}\relax
\mciteBstWouldAddEndPuncttrue
\mciteSetBstMidEndSepPunct{\mcitedefaultmidpunct}
{\mcitedefaultendpunct}{\mcitedefaultseppunct}\relax
\EndOfBibitem
\bibitem{Agaev:2017lip}
S.~S. Agaev, K.~Azizi, and H.~Sundu,
  \ifthenelse{\boolean{articletitles}}{\emph{{Interpretation of the new
  $\POmega_c^{0}$ states via their mass and width}},
  }{}\href{https://doi.org/10.1140/epjc/s10052-017-4953-z}{Eur.\ Phys.\ J.\
  \textbf{C77} (2017) 395},
  \href{http://arxiv.org/abs/1704.04928}{{\normalfont\ttfamily
  arXiv:1704.04928}}\relax
\mciteBstWouldAddEndPuncttrue
\mciteSetBstMidEndSepPunct{\mcitedefaultmidpunct}
{\mcitedefaultendpunct}{\mcitedefaultseppunct}\relax
\EndOfBibitem
\bibitem{Chen:2017sci}
H.-X. Chen {\em et~al.}, \ifthenelse{\boolean{articletitles}}{\emph{{Decay
  properties of $P$-wave charmed baryons from light-cone QCD sum rules}},
  }{}\href{https://doi.org/10.1103/PhysRevD.95.094008}{Phys.\ Rev.\
  \textbf{D95} (2017) 094008},
  \href{http://arxiv.org/abs/1703.07703}{{\normalfont\ttfamily
  arXiv:1703.07703}}\relax
\mciteBstWouldAddEndPuncttrue
\mciteSetBstMidEndSepPunct{\mcitedefaultmidpunct}
{\mcitedefaultendpunct}{\mcitedefaultseppunct}\relax
\EndOfBibitem
\bibitem{Chen:2015kpa}
H.-X. Chen {\em et~al.}, \ifthenelse{\boolean{articletitles}}{\emph{{P-wave
  charmed baryons from QCD sum rules}},
  }{}\href{https://doi.org/10.1103/PhysRevD.91.054034}{Phys.\ Rev.\
  \textbf{D91} (2015) 054034},
  \href{http://arxiv.org/abs/1502.01103}{{\normalfont\ttfamily
  arXiv:1502.01103}}\relax
\mciteBstWouldAddEndPuncttrue
\mciteSetBstMidEndSepPunct{\mcitedefaultmidpunct}
{\mcitedefaultendpunct}{\mcitedefaultseppunct}\relax
\EndOfBibitem
\bibitem{Wang:2017zjw}
Z.-G. Wang, \ifthenelse{\boolean{articletitles}}{\emph{{Analysis of $\POmega
  _\cquark(3000)$, $\POmega _\cquark(3050)$, $\POmega _\cquark(3066)$, $\POmega
  _\cquark(3090)$ and $\POmega _\cquark(3119)$ with QCD sum rules}},
  }{}\href{https://doi.org/10.1140/epjc/s10052-017-4895-5}{Eur.\ Phys.\ J.\
  \textbf{C77} (2017) 325},
  \href{http://arxiv.org/abs/1704.01854}{{\normalfont\ttfamily
  arXiv:1704.01854}}\relax
\mciteBstWouldAddEndPuncttrue
\mciteSetBstMidEndSepPunct{\mcitedefaultmidpunct}
{\mcitedefaultendpunct}{\mcitedefaultseppunct}\relax
\EndOfBibitem
\bibitem{Chen:2017xat}
R.~Chen, A.~Hosaka, and X.~Liu,
  \ifthenelse{\boolean{articletitles}}{\emph{{Searching for possible
  $\POmega_\cquark$-like molecular states from meson-baryon interaction}},
  }{}\href{https://doi.org/10.1103/PhysRevD.97.036016}{Phys.\ Rev.\
  \textbf{D97} (2018) 036016},
  \href{http://arxiv.org/abs/1711.07650}{{\normalfont\ttfamily
  arXiv:1711.07650}}\relax
\mciteBstWouldAddEndPuncttrue
\mciteSetBstMidEndSepPunct{\mcitedefaultmidpunct}
{\mcitedefaultendpunct}{\mcitedefaultseppunct}\relax
\EndOfBibitem
\bibitem{Kim:2017jpx}
H.-C. Kim, M.~V. Polyakov, and M.~Prasza\l{}owicz,
  \ifthenelse{\boolean{articletitles}}{\emph{{Possibility of the existence of
  charmed exotica}},
  }{}\href{https://doi.org/10.1103/PhysRevD.96.014009}{Phys.\ Rev.\
  \textbf{D96} (2017) 014009}, Erratum
  \href{https://doi.org/10.1103/PhysRevD.96.039902}{ibid.\   \textbf{D96}
  (2017) 039902}, \href{http://arxiv.org/abs/1704.04082}{{\normalfont\ttfamily
  arXiv:1704.04082}}\relax
\mciteBstWouldAddEndPuncttrue
\mciteSetBstMidEndSepPunct{\mcitedefaultmidpunct}
{\mcitedefaultendpunct}{\mcitedefaultseppunct}\relax
\EndOfBibitem
\bibitem{An:2017lwg}
C.~S. An and H.~Chen, \ifthenelse{\boolean{articletitles}}{\emph{{Observed
  $\POmega_{\cquark}^{0}$ resonances as pentaquark states}},
  }{}\href{https://doi.org/10.1103/PhysRevD.96.034012}{Phys.\ Rev.\
  \textbf{D96} (2017) 034012},
  \href{http://arxiv.org/abs/1705.08571}{{\normalfont\ttfamily
  arXiv:1705.08571}}\relax
\mciteBstWouldAddEndPuncttrue
\mciteSetBstMidEndSepPunct{\mcitedefaultmidpunct}
{\mcitedefaultendpunct}{\mcitedefaultseppunct}\relax
\EndOfBibitem
\bibitem{Ali:2017wsf}
A.~Ali {\em et~al.}, \ifthenelse{\boolean{articletitles}}{\emph{{A new look at
  the Y tetraquarks and $\POmega _\cquark$ baryons in the diquark model}},
  }{}\href{https://doi.org/10.1140/epjc/s10052-017-5501-6}{Eur.\ Phys.\ J.\
  \textbf{C78} (2018) 29},
  \href{http://arxiv.org/abs/1708.04650}{{\normalfont\ttfamily
  arXiv:1708.04650}}\relax
\mciteBstWouldAddEndPuncttrue
\mciteSetBstMidEndSepPunct{\mcitedefaultmidpunct}
{\mcitedefaultendpunct}{\mcitedefaultseppunct}\relax
\EndOfBibitem
\bibitem{Montana:2017kjw}
G.~Monta\~na, A.~Feijoo, and A.~Ramos,
  \ifthenelse{\boolean{articletitles}}{\emph{{A meson-baryon molecular
  interpretation for some $\POmega_{\cquark}$ excited states}},
  }{}\href{https://doi.org/10.1140/epja/i2018-12498-1}{Eur.\ Phys.\ J.\
  \textbf{A54} (2018) 64},
  \href{http://arxiv.org/abs/1709.08737}{{\normalfont\ttfamily
  arXiv:1709.08737}}\relax
\mciteBstWouldAddEndPuncttrue
\mciteSetBstMidEndSepPunct{\mcitedefaultmidpunct}
{\mcitedefaultendpunct}{\mcitedefaultseppunct}\relax
\EndOfBibitem
\bibitem{Debastiani:2017ewu}
V.~R. Debastiani, J.~M. Dias, W.~H. Liang, and E.~Oset,
  \ifthenelse{\boolean{articletitles}}{\emph{{Molecular $\Omegares_\cquark$
  states generated from coupled meson-baryon channels}},
  }{}\href{https://doi.org/10.1103/PhysRevD.97.094035}{Phys.\ Rev.\
  \textbf{D97} (2018) 094035},
  \href{http://arxiv.org/abs/1710.04231}{{\normalfont\ttfamily
  arXiv:1710.04231}}\relax
\mciteBstWouldAddEndPuncttrue
\mciteSetBstMidEndSepPunct{\mcitedefaultmidpunct}
{\mcitedefaultendpunct}{\mcitedefaultseppunct}\relax
\EndOfBibitem
\bibitem{Santopinto:2018ljf}
E.~Santopinto {\em et~al.}, \ifthenelse{\boolean{articletitles}}{\emph{{The
  $\varOmega _{ c}$-puzzle solved by means of quark model predictions}},
  }{}\href{https://doi.org/10.1140/epjc/s10052-019-7527-4}{Eur.\ Phys.\ J.\ C
  \textbf{79} (2019) 1012},
  \href{http://arxiv.org/abs/1811.01799}{{\normalfont\ttfamily
  arXiv:1811.01799}}\relax
\mciteBstWouldAddEndPuncttrue
\mciteSetBstMidEndSepPunct{\mcitedefaultmidpunct}
{\mcitedefaultendpunct}{\mcitedefaultseppunct}\relax
\EndOfBibitem
\bibitem{Debastiani:2018adr}
V.~R. Debastiani, J.~M. Dias, W.-H. Liang, and E.~Oset,
  \ifthenelse{\boolean{articletitles}}{\emph{{$\POmega_\bquark^- \to
  (\PXi_\cquark^+ \, K^-) \, \pi^-$ and the $\POmega_\cquark$ states}},
  }{}\href{https://doi.org/10.1103/PhysRevD.98.094022}{Phys.\ Rev.\
  \textbf{D98} (2018) 094022},
  \href{http://arxiv.org/abs/1803.03268}{{\normalfont\ttfamily
  arXiv:1803.03268}}\relax
\mciteBstWouldAddEndPuncttrue
\mciteSetBstMidEndSepPunct{\mcitedefaultmidpunct}
{\mcitedefaultendpunct}{\mcitedefaultseppunct}\relax
\EndOfBibitem
\bibitem{Chua:2019yqh}
C.-K. Chua, \ifthenelse{\boolean{articletitles}}{\emph{{Color-allowed bottom
  baryon to $s$-wave and $p$-wave charmed baryon nonleptonic decays}},
  }{}\href{https://doi.org/10.1103/PhysRevD.100.034025}{Phys.\ Rev.\
  \textbf{D100} (2019) 034025},
  \href{http://arxiv.org/abs/1905.00153}{{\normalfont\ttfamily
  arXiv:1905.00153}}\relax
\mciteBstWouldAddEndPuncttrue
\mciteSetBstMidEndSepPunct{\mcitedefaultmidpunct}
{\mcitedefaultendpunct}{\mcitedefaultseppunct}\relax
\EndOfBibitem
\bibitem{LHCb-DP-2008-001}
LHCb collaboration, A.~A. Alves~Jr.\ {\em et~al.},
  \ifthenelse{\boolean{articletitles}}{\emph{{The \lhcb detector at the LHC}},
  }{}\href{https://doi.org/10.1088/1748-0221/3/08/S08005}{JINST \textbf{3}
  (2008) S08005}\relax
\mciteBstWouldAddEndPuncttrue
\mciteSetBstMidEndSepPunct{\mcitedefaultmidpunct}
{\mcitedefaultendpunct}{\mcitedefaultseppunct}\relax
\EndOfBibitem
\bibitem{LHCb-DP-2014-002}
LHCb collaboration, R.~Aaij {\em et~al.},
  \ifthenelse{\boolean{articletitles}}{\emph{{LHCb detector performance}},
  }{}\href{https://doi.org/10.1142/S0217751X15300227}{Int.\ J.\ Mod.\ Phys.\
  \textbf{A30} (2015) 1530022},
  \href{http://arxiv.org/abs/1412.6352}{{\normalfont\ttfamily
  arXiv:1412.6352}}\relax
\mciteBstWouldAddEndPuncttrue
\mciteSetBstMidEndSepPunct{\mcitedefaultmidpunct}
{\mcitedefaultendpunct}{\mcitedefaultseppunct}\relax
\EndOfBibitem
\bibitem{Sjostrand:2007gs}
T.~Sj\"{o}strand, S.~Mrenna, and P.~Skands,
  \ifthenelse{\boolean{articletitles}}{\emph{{A brief introduction to PYTHIA
  8.1}}, }{}\href{https://doi.org/10.1016/j.cpc.2008.01.036}{Comput.\ Phys.\
  Commun.\  \textbf{178} (2008) 852},
  \href{http://arxiv.org/abs/0710.3820}{{\normalfont\ttfamily
  arXiv:0710.3820}}\relax
\mciteBstWouldAddEndPuncttrue
\mciteSetBstMidEndSepPunct{\mcitedefaultmidpunct}
{\mcitedefaultendpunct}{\mcitedefaultseppunct}\relax
\EndOfBibitem
\bibitem{Sjostrand:2006za}
T.~Sj\"{o}strand, S.~Mrenna, and P.~Skands,
  \ifthenelse{\boolean{articletitles}}{\emph{{PYTHIA 6.4 physics and manual}},
  }{}\href{https://doi.org/10.1088/1126-6708/2006/05/026}{JHEP \textbf{05}
  (2006) 026}, \href{http://arxiv.org/abs/hep-ph/0603175}{{\normalfont\ttfamily
  arXiv:hep-ph/0603175}}\relax
\mciteBstWouldAddEndPuncttrue
\mciteSetBstMidEndSepPunct{\mcitedefaultmidpunct}
{\mcitedefaultendpunct}{\mcitedefaultseppunct}\relax
\EndOfBibitem
\bibitem{LHCb-PROC-2010-056}
I.~Belyaev {\em et~al.}, \ifthenelse{\boolean{articletitles}}{\emph{{Handling
  of the generation of primary events in Gauss, the LHCb simulation
  framework}}, }{}\href{https://doi.org/10.1088/1742-6596/331/3/032047}{J.\
  Phys.\ Conf.\ Ser.\  \textbf{331} (2011) 032047}\relax
\mciteBstWouldAddEndPuncttrue
\mciteSetBstMidEndSepPunct{\mcitedefaultmidpunct}
{\mcitedefaultendpunct}{\mcitedefaultseppunct}\relax
\EndOfBibitem
\bibitem{Lange:2001uf}
D.~J. Lange, \ifthenelse{\boolean{articletitles}}{\emph{{The EvtGen particle
  decay simulation package}},
  }{}\href{https://doi.org/10.1016/S0168-9002(01)00089-4}{Nucl.\ Instrum.\
  Meth.\  \textbf{A462} (2001) 152}\relax
\mciteBstWouldAddEndPuncttrue
\mciteSetBstMidEndSepPunct{\mcitedefaultmidpunct}
{\mcitedefaultendpunct}{\mcitedefaultseppunct}\relax
\EndOfBibitem
\bibitem{Allison:2006ve}
Geant4 collaboration, J.~Allison {\em et~al.},
  \ifthenelse{\boolean{articletitles}}{\emph{{Geant4 developments and
  applications}}, }{}\href{https://doi.org/10.1109/TNS.2006.869826}{IEEE
  Trans.\ Nucl.\ Sci.\  \textbf{53} (2006) 270}\relax
\mciteBstWouldAddEndPuncttrue
\mciteSetBstMidEndSepPunct{\mcitedefaultmidpunct}
{\mcitedefaultendpunct}{\mcitedefaultseppunct}\relax
\EndOfBibitem
\bibitem{Agostinelli:2002hh}
Geant4 collaboration, S.~Agostinelli {\em et~al.},
  \ifthenelse{\boolean{articletitles}}{\emph{{Geant4: A simulation toolkit}},
  }{}\href{https://doi.org/10.1016/S0168-9002(03)01368-8}{Nucl.\ Instrum.\
  Meth.\  \textbf{A506} (2003) 250}\relax
\mciteBstWouldAddEndPuncttrue
\mciteSetBstMidEndSepPunct{\mcitedefaultmidpunct}
{\mcitedefaultendpunct}{\mcitedefaultseppunct}\relax
\EndOfBibitem
\bibitem{LHCb-PROC-2011-006}
M.~Clemencic {\em et~al.}, \ifthenelse{\boolean{articletitles}}{\emph{{The
  \lhcb simulation application, Gauss: Design, evolution and experience}},
  }{}\href{https://doi.org/10.1088/1742-6596/331/3/032023}{J.\ Phys.\ Conf.\
  Ser.\  \textbf{331} (2011) 032023}\relax
\mciteBstWouldAddEndPuncttrue
\mciteSetBstMidEndSepPunct{\mcitedefaultmidpunct}
{\mcitedefaultendpunct}{\mcitedefaultseppunct}\relax
\EndOfBibitem
\bibitem{Hocker:2007ht}
H.~Voss, A.~Hoecker, J.~Stelzer, and F.~Tegenfeldt,
  \ifthenelse{\boolean{articletitles}}{\emph{{TMVA - Toolkit for Multivariate
  Data Analysis with ROOT}}, }{}\href{https://doi.org/10.22323/1.050.0040}{PoS
  \textbf{ACAT} (2007) 040}\relax
\mciteBstWouldAddEndPuncttrue
\mciteSetBstMidEndSepPunct{\mcitedefaultmidpunct}
{\mcitedefaultendpunct}{\mcitedefaultseppunct}\relax
\EndOfBibitem
\bibitem{TMVA4}
A.~Hoecker {\em et~al.}, \ifthenelse{\boolean{articletitles}}{\emph{{TMVA 4 ---
  Toolkit for Multivariate Data Analysis with ROOT. Users Guide.}},
  }{}\href{http://arxiv.org/abs/physics/0703039}{{\normalfont\ttfamily
  arXiv:physics/0703039}}\relax
\mciteBstWouldAddEndPuncttrue
\mciteSetBstMidEndSepPunct{\mcitedefaultmidpunct}
{\mcitedefaultendpunct}{\mcitedefaultseppunct}\relax
\EndOfBibitem
\bibitem{Punzi:2003bu}
G.~Punzi, \ifthenelse{\boolean{articletitles}}{\emph{{Sensitivity of searches
  for new signals and its optimization}}, }{}eConf \textbf{C030908} (2003)
  MODT002, \href{http://arxiv.org/abs/physics/0308063}{{\normalfont\ttfamily
  arXiv:physics/0308063}}\relax
\mciteBstWouldAddEndPuncttrue
\mciteSetBstMidEndSepPunct{\mcitedefaultmidpunct}
{\mcitedefaultendpunct}{\mcitedefaultseppunct}\relax
\EndOfBibitem
\bibitem{Hulsbergen:2005pu}
W.~D. Hulsbergen, \ifthenelse{\boolean{articletitles}}{\emph{{Decay chain
  fitting with a Kalman filter}},
  }{}\href{https://doi.org/10.1016/j.nima.2005.06.078}{Nucl.\ Instrum.\ Meth.\
  \textbf{A552} (2005) 566},
  \href{http://arxiv.org/abs/physics/0503191}{{\normalfont\ttfamily
  arXiv:physics/0503191}}\relax
\mciteBstWouldAddEndPuncttrue
\mciteSetBstMidEndSepPunct{\mcitedefaultmidpunct}
{\mcitedefaultendpunct}{\mcitedefaultseppunct}\relax
\EndOfBibitem
\bibitem{PDG2020}
Particle Data Group, P.~A. Zyla {\em et~al.},
  \ifthenelse{\boolean{articletitles}}{\emph{{\href{http://pdg.lbl.gov/}{Review
  of particle physics}}}, }{}\href{https://doi.org/10.1093/ptep/ptaa104}{Prog.\
  Theor.\ Exp.\ Phys.\  \textbf{2020} (2020) 083C01}\relax
\mciteBstWouldAddEndPuncttrue
\mciteSetBstMidEndSepPunct{\mcitedefaultmidpunct}
{\mcitedefaultendpunct}{\mcitedefaultseppunct}\relax
\EndOfBibitem
\bibitem{Cowan:2016tnm}
G.~A. Cowan, D.~C. Craik, and M.~D. Needham,
  \ifthenelse{\boolean{articletitles}}{\emph{{RapidSim: an application for the
  fast simulation of heavy-quark hadron decays}},
  }{}\href{https://doi.org/10.1016/j.cpc.2017.01.029}{Comput.\ Phys.\ Commun.\
  \textbf{214} (2017) 239},
  \href{http://arxiv.org/abs/1612.07489}{{\normalfont\ttfamily
  arXiv:1612.07489}}\relax
\mciteBstWouldAddEndPuncttrue
\mciteSetBstMidEndSepPunct{\mcitedefaultmidpunct}
{\mcitedefaultendpunct}{\mcitedefaultseppunct}\relax
\EndOfBibitem
\bibitem{DeMaesschalck:2000xyz}
R.~{De Maesschalck}, D.~Jouan-Rimbaud, and D.~L. Massart,
  \ifthenelse{\boolean{articletitles}}{\emph{{The Mahalanobis distance}},
  }{}\href{https://doi.org/https://doi.org/10.1016/S0169-7439(99)00047-7}{{Chemometrics
  and Intelligent Laboratory Systems} \textbf{50} (2000) 1}\relax
\mciteBstWouldAddEndPuncttrue
\mciteSetBstMidEndSepPunct{\mcitedefaultmidpunct}
{\mcitedefaultendpunct}{\mcitedefaultseppunct}\relax
\EndOfBibitem
\bibitem{LHCb-PAPER-2021-012-supp}
See Supplemental Material at
  \url{http://link.aps.org/supplemental/10.1103/PhysRevD.104.L091102} for
  tables of systematic uncertainties and additional plots.\relax
\mciteBstWouldAddEndPunctfalse
\mciteSetBstMidEndSepPunct{\mcitedefaultmidpunct}
{}{\mcitedefaultseppunct}\relax
\EndOfBibitem
\bibitem{LHCb-PAPER-2013-011}
LHCb collaboration, R.~Aaij {\em et~al.},
  \ifthenelse{\boolean{articletitles}}{\emph{{Precision measurement of \D meson
  mass differences}}, }{}\href{https://doi.org/10.1007/JHEP06(2013)065}{JHEP
  \textbf{06} (2013) 065},
  \href{http://arxiv.org/abs/1304.6865}{{\normalfont\ttfamily
  arXiv:1304.6865}}\relax
\mciteBstWouldAddEndPuncttrue
\mciteSetBstMidEndSepPunct{\mcitedefaultmidpunct}
{\mcitedefaultendpunct}{\mcitedefaultseppunct}\relax
\EndOfBibitem
\bibitem{Blatt:1952ije}
J.~M. Blatt and V.~F. Weisskopf, {\em {Theoretical nuclear physics}},
  \href{https://doi.org/10.1007/978-1-4612-9959-2}{ Springer, New York,
  1952}\relax
\mciteBstWouldAddEndPuncttrue
\mciteSetBstMidEndSepPunct{\mcitedefaultmidpunct}
{\mcitedefaultendpunct}{\mcitedefaultseppunct}\relax
\EndOfBibitem
\bibitem{VonHippel:1972fg}
F.~Von~Hippel and C.~Quigg,
  \ifthenelse{\boolean{articletitles}}{\emph{{Centrifugal-barrier effects in
  resonance partial decay widths, shapes, and production amplitudes}},
  }{}\href{https://doi.org/10.1103/PhysRevD.5.624}{Phys.\ Rev.\  \textbf{D5}
  (1972) 624}\relax
\mciteBstWouldAddEndPuncttrue
\mciteSetBstMidEndSepPunct{\mcitedefaultmidpunct}
{\mcitedefaultendpunct}{\mcitedefaultseppunct}\relax
\EndOfBibitem
\bibitem{LHCb-PAPER-2016-008}
LHCb collaboration, R.~Aaij {\em et~al.},
  \ifthenelse{\boolean{articletitles}}{\emph{{Measurements of the mass and
  lifetime of the \Omegab baryon}},
  }{}\href{https://doi.org/10.1103/PhysRevD.93.092007}{Phys.\ Rev.\
  \textbf{D93} (2016) 092007},
  \href{http://arxiv.org/abs/1604.01412}{{\normalfont\ttfamily
  arXiv:1604.01412}}\relax
\mciteBstWouldAddEndPuncttrue
\mciteSetBstMidEndSepPunct{\mcitedefaultmidpunct}
{\mcitedefaultendpunct}{\mcitedefaultseppunct}\relax
\EndOfBibitem
\bibitem{LHCb-PAPER-2012-048}
LHCb collaboration, R.~Aaij {\em et~al.},
  \ifthenelse{\boolean{articletitles}}{\emph{{Measurements of the \Lb, \Xibm,
  and \Omegab baryon masses}},
  }{}\href{https://doi.org/10.1103/PhysRevLett.110.182001}{Phys.\ Rev.\ Lett.\
  \textbf{110} (2013) 182001},
  \href{http://arxiv.org/abs/1302.1072}{{\normalfont\ttfamily
  arXiv:1302.1072}}\relax
\mciteBstWouldAddEndPuncttrue
\mciteSetBstMidEndSepPunct{\mcitedefaultmidpunct}
{\mcitedefaultendpunct}{\mcitedefaultseppunct}\relax
\EndOfBibitem
\end{mcitethebibliography}

\clearpage

\section*{Supplemental material for LHCb-PAPER-2021-012}
\label{sec:Supplementary-App}

\begin{figure}[h]
    \centering
    \includegraphics[width=0.49\textwidth]{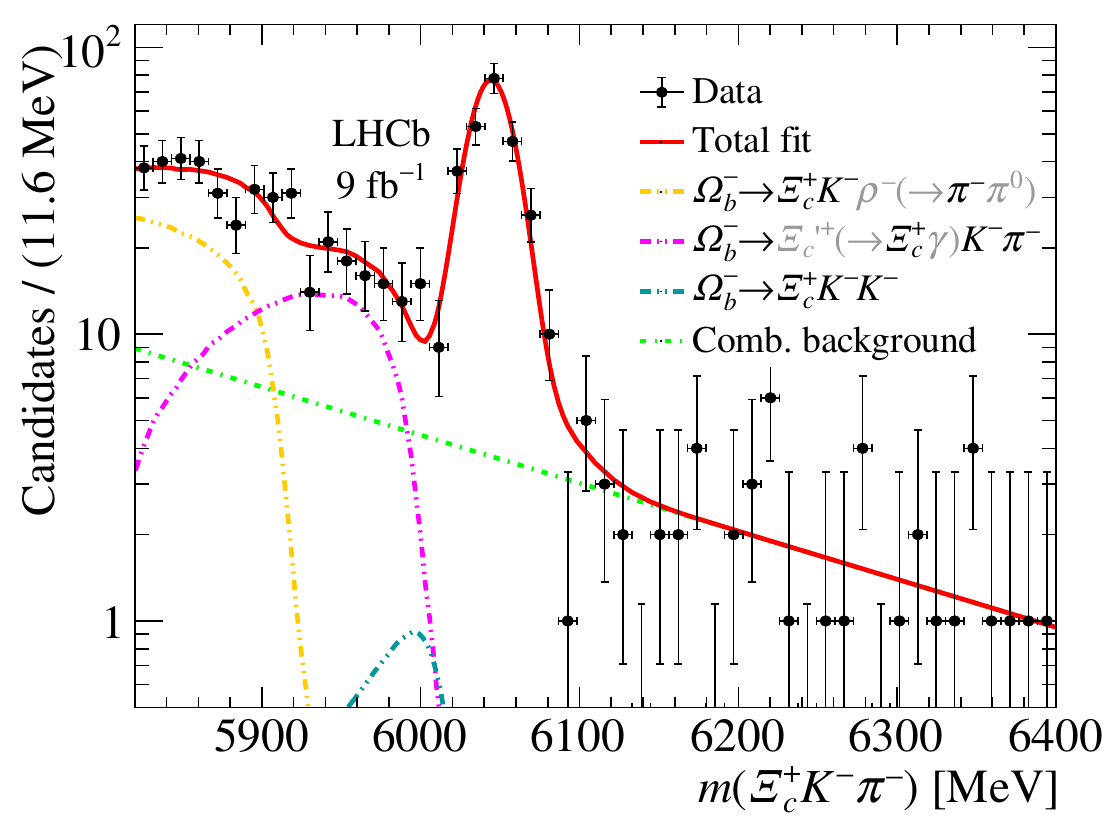}
    \includegraphics[width=0.49\textwidth]{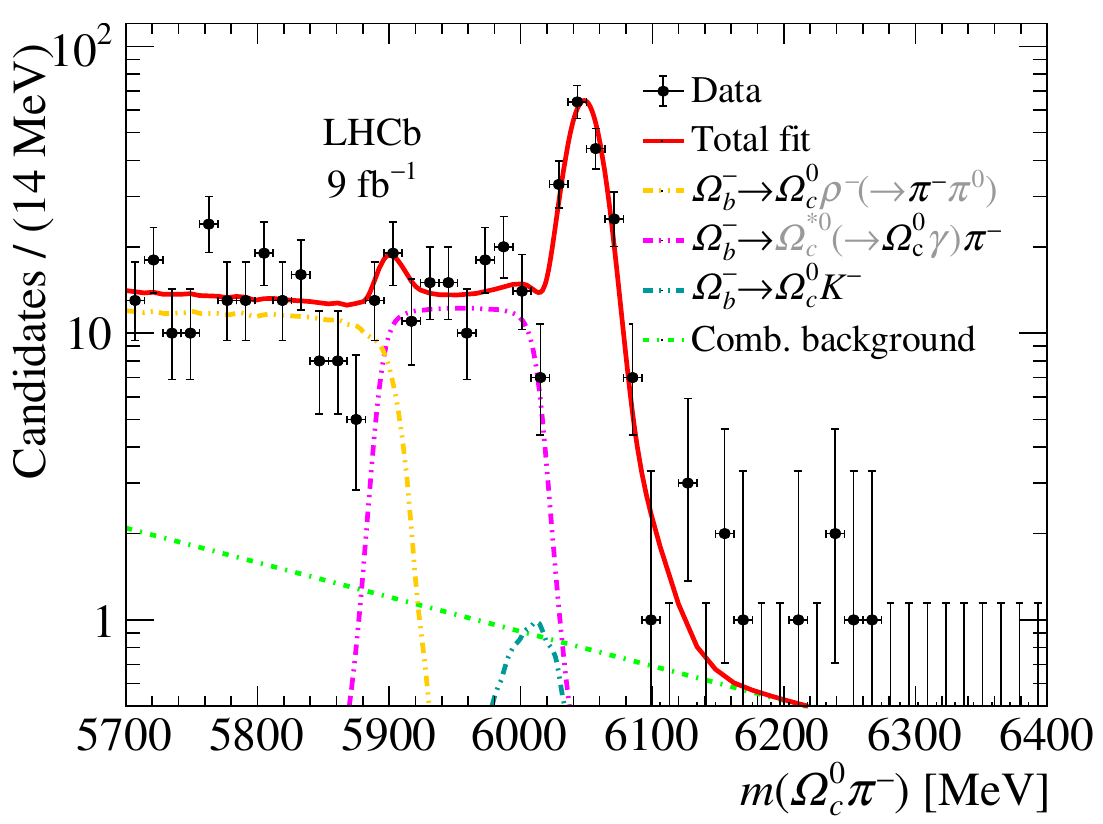}
    \caption{Distribution of the reconstructed invariant mass (left) $m(\Xicp\Km\pim)$ with $\Xicp \to \proton\Km\pip$ and (right) $m(\Omegac\pim)$ with $\Omegac \to \proton\Km\Km\pip$ using a logarithmic scale for all candidates passing the selection requirements. The black symbols show the selected signal candidates. The result of a fit is overlaid (solid red line). 
    The missing particles in partially reconstructed decays are indicated in grey in the legends.}
    \label{fig:ObFitLog}
\end{figure}

The mass distributions, $m(\Xicp\Km\pim)$ and $m(\Omegac\pim)$, are shown in Fig.~\ref{fig:ObFitLog} with a logarithmic scale. 

\begin{figure}[h]
    \centering
    \includegraphics[width = 0.7\textwidth]{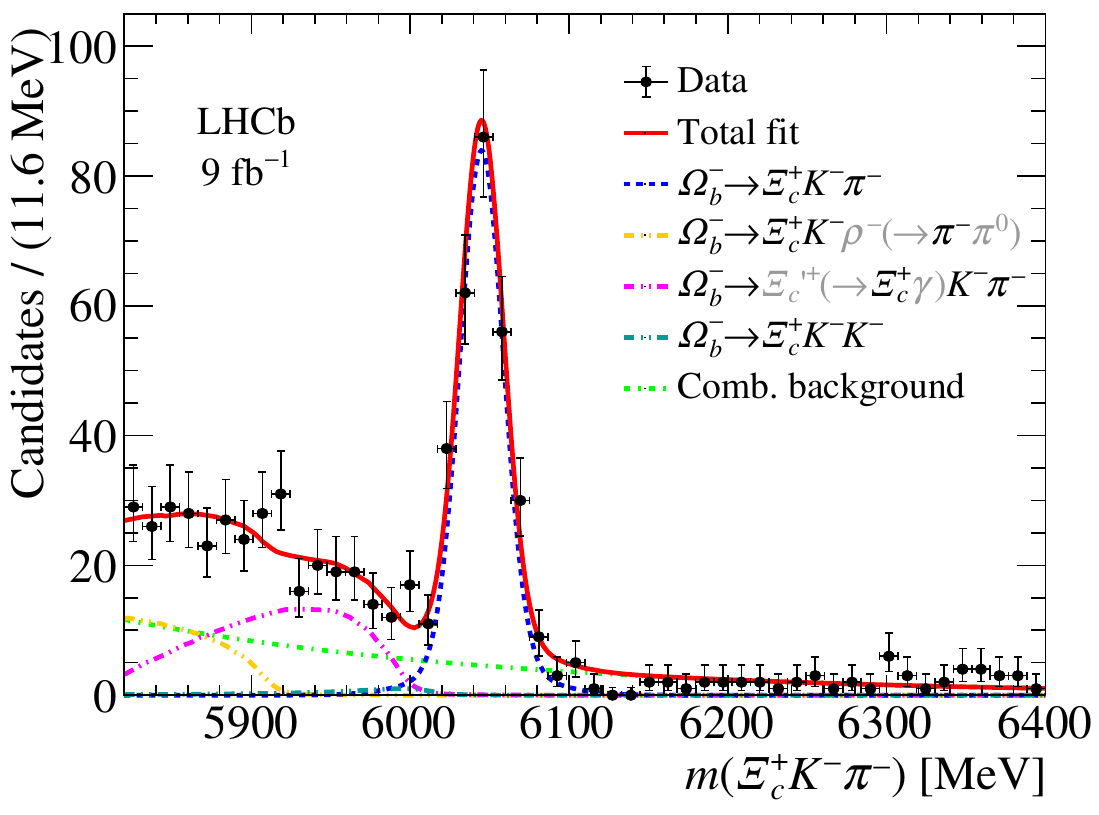}
    \caption{Distribution of the reconstructed invariant mass $m(\Xicp\Km\pim)$ with $\Xicp \to \proton\Km\pip$, where the simulation used to train the BDT classifier has a requirement of $m(\Xicp\Km) < 3.3\gev$. }
    \label{fig:ObFitXicK}
\end{figure}

\begin{figure}
    \centering
    \includegraphics[width=0.7\textwidth]{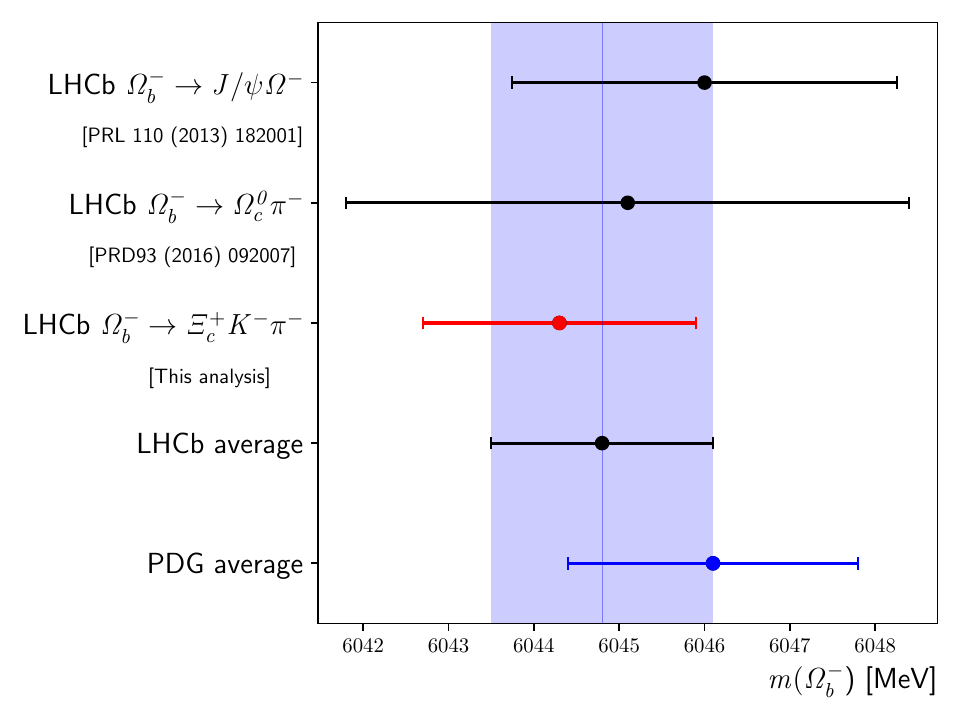}
    \caption{Measurements of the \Omegab mass from the \lhcb experiment, the \lhcb average and the PDG average, which includes the two previous \lhcb measurements and one measurement from the CDF collaboration~\cite{PDG2020}.} 
    \label{fig:AveragedMass}
\end{figure}

The \Xicp\Km\pim mass distribution used for the investigation of the \OmegacXX states is shown in Fig.~\ref{fig:ObFitXicK}, where a new BDT classifier is trained with the addition of the requirement $m(\Xicp\Km) < 3.3\gev$ in the simulation. 
Figure~\ref{fig:AveragedMass} shows all measurements of the \Omegab mass from the LHCb experiment, the LHCb average which is calculated using the \Omegab mass determined in this analysis and the two previous results, and the PDG average.

\begin{figure}
    \centering
    \includegraphics[width=0.9\linewidth]{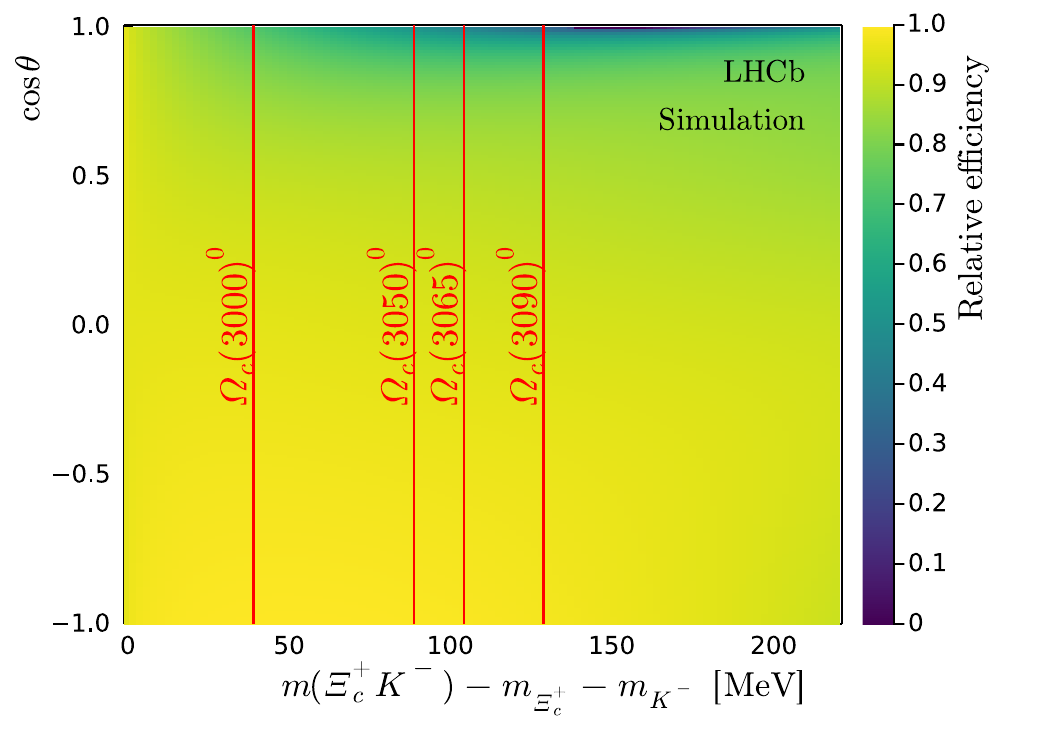}
    \caption{Efficiency map in the $[\Delta M,\, \cos\theta]$ plane.
    Positions of the four narrow $\OmegacXX$ states are shown by the red lines.}
    \label{fig:efficiency}
\end{figure}

\begin{figure}[t]
    \centering
    \includegraphics[width=0.48\textwidth]{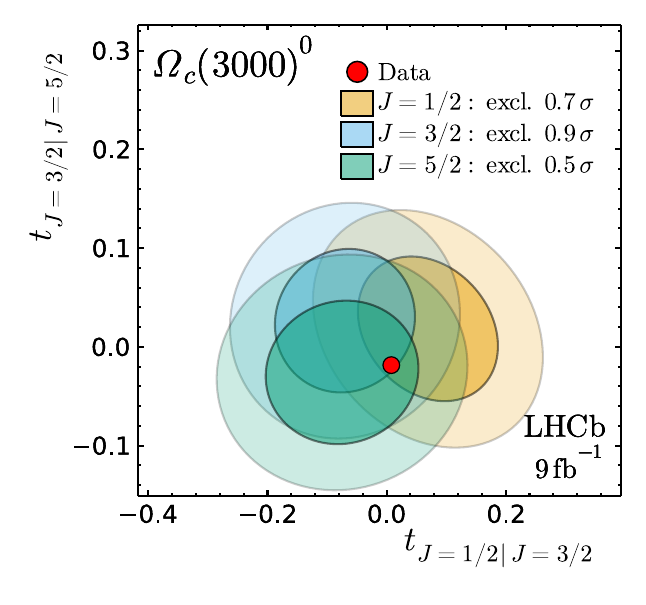}
    \includegraphics[width=0.48\textwidth]{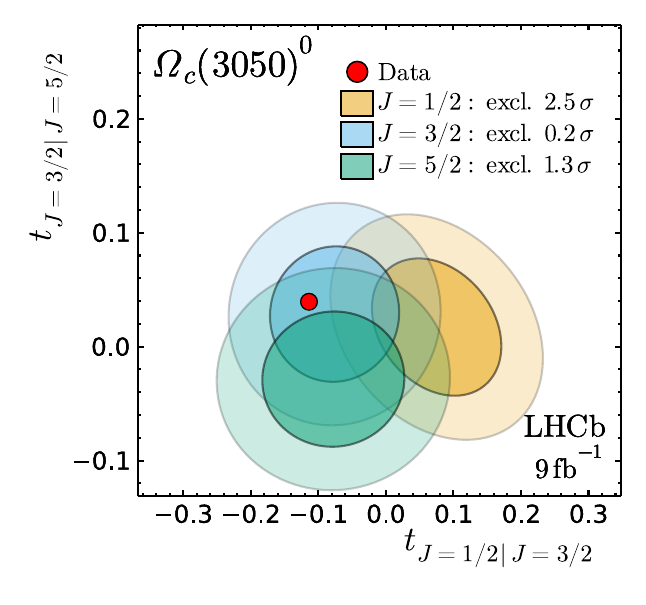}
    \includegraphics[width=0.48\textwidth]{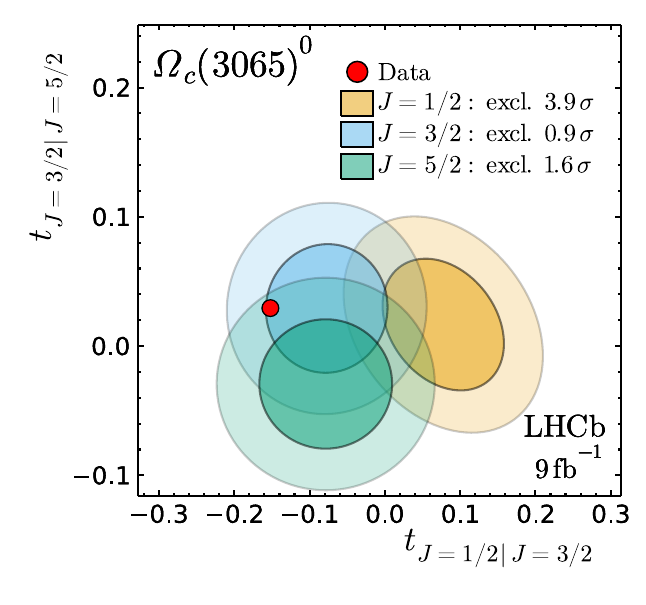}
    \includegraphics[width=0.48\textwidth]{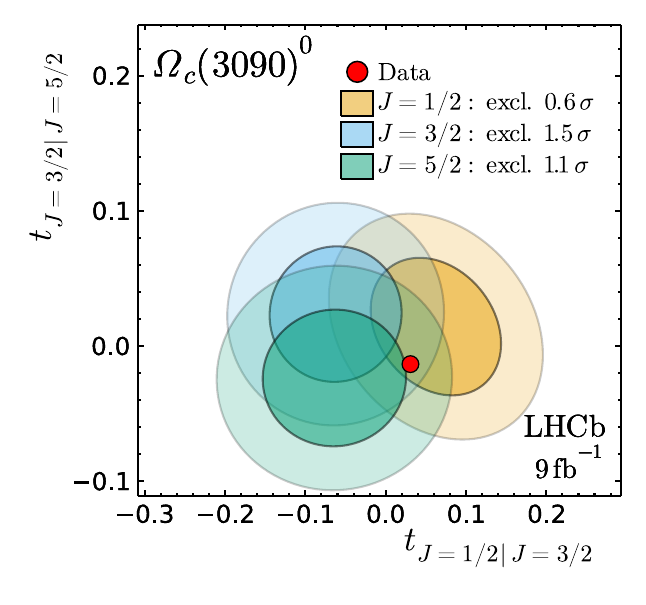}
    \caption{
    Values of the spin-hypothesis estimators $t_{J=1/2|J=3/2}$ and $t_{J=3/2|J=5/2}$.
    The red point shows the value measured in the default fit to the data.
    The colored ellipses give the $67\,\%$ and $95\,\%$
    confidence intervals in the probability density for spin hypotheses with $J=1/2$, $3/2$, and $5/2$ as indicated in the legend.
    The rejection significance of every spin-$J$ hypothesis are shown in the legend for the default fit.
    The values with systematic studies included are summarised in the Table~\ref{tab:final_results} of the main text.
    }
    \label{fig:hypotheses.test}
\end{figure}

The efficiency map of the data reconstruction and selection in the $[\Delta M,\,\cos\theta]$ plane is shown in Fig.~\ref{fig:efficiency}, where the positions of four of the \OmegacXX states are demonstrated by the red lines. 
Figure~\ref{fig:hypotheses.test} shows the value of the test statistic observables $t_{J=1/2|J=3/2}$ and $t_{J=3/2|J=5/2}$ for each \OmegacXX state, where the red point indicates the value determined from data. 
\clearpage
Tables~\ref{tab:syst} and~\ref{tab:XicKsyst} summarise the systematic uncertainties considered for each observable where the largest deviation from the default model is quoted on every source. 

\begin{table}
    \caption{Systematic uncertainties in the \Omegab mass,
    relative branching fraction \pcalXiKpioverOcpi,
    and relative production rate \pcal{} of the \OmegacXX states.
    }
    \label{tab:syst}
    \centering
    \resizebox{0.99\columnwidth}{!}{
    \begin{tabular}{l|c c | c c | c c}
                & \multicolumn{2}{c|}{\Omegab} & \OmegacXXa & \OmegacXXb & \OmegacXXc & \OmegacXXd\\
         Source & $m$ [\mev\kern -0.2em] & \pcalXiKpioverOcpi & \pcal{} & \pcal{} & \pcal{} & \pcal{}\\
         \hline
         Alternative \Omegab fit & $<0.1$ & $0.05$ & $<0.01$ & $<0.01$ & $<0.01$ & $<0.01$ \\
         \Xicp Dalitz weights & $0.5$ & $<0.01$ & $0.03$ & $0.01$ & $<0.01$ & $0.02$ \\
         Momentum calibration &  $0.9$ & --- & --- & --- & --- & --- \\
         PID efficiency & $0.2$ & $0.01$ & $<0.01$ & $0.02$ & $0.01$ & $0.03$ \\
         \Omegab kinematics & $0.4$ & $<0.01$ & $0.01$ & $<0.01$ & $0.01$ & $0.01$ \\
         Alternative \Xicp\Km fit & --- & --- &  $0.02$ & $0.01$ & $0.02$ & $0.01$ \\
         Efficiency map & --- & $<0.01$ & $<0.01$ & $<0.01$ & $<0.01$ & $<0.01$\\
         Background in \Xicp\Km & --- & --- & $<0.01$ & $<0.01$ & $<0.01$ & $<0.01$ \\
         \hline
         Total & $1.1$ & $0.05$ & $0.04$ & $0.02$ & $0.02$ & $0.04$ \\
         \hline
         
    \end{tabular}
    }
\end{table}

\begin{table}[]
    \centering
    \caption{Systematic uncertainties in the resonance parameters of the \OmegacXX states.
    For the width of the \OmegacXXb baryon, the systematic uncertainties are embedded in the upper limit.
    }
    \label{tab:XicKsyst}
    \resizebox{\columnwidth}{!}{
    \begin{tabular}{l|c c | c | c c | c c}
         & \multicolumn{2}{c|}{\OmegacXXa} & \OmegacXXb & \multicolumn{2}{c|}{\OmegacXXc} & \multicolumn{2}{c}{\OmegacXXd}\\
         Source & $m$ [\mev\kern -0.2em] & $\Gamma$ [\mev\kern -0.2em] & $m$ [\mev\kern -0.2em]  & $m$ [\mev\kern -0.2em] & $\Gamma$ [\mev\kern -0.2em] & $m$ [\mev\kern -0.2em] & $\Gamma$ [\mev\kern -0.2em] \\
         \hline
        Alternative \Omegab fit  & $<0.01$ & $<0.01$ & $< 0.01$ & $<0.01$ & $<0.01$ & $<0.01$ & $<0.01$ \\
        \Xicp Dalitz weights     & $ 0.02$ & $ 1.1 $ & $0.10$ & $0.14$ & $0.2$ & $0.32$ & $1.2$ \\
        Momentum calibration     & $ 0.01$ &  ---    & $0.03$ & $0.03$ & --- & $0.04$ & --- \\
        PID efficiency           & $ 0.56$ & $ 0.1$ & $0.05$ & $0.14$ & $0.2$ & $0.73$ & $2.1$ \\
        \Omegab kinematics       & $ 0.13$ & $ 0.7$ & $0.10$ & $ 0.21$ & $0.2$ & $ 0.42$ & $ 0.9$ \\
        Alternative \Xicp\Km fit & $ 0.70$ & $2.1$ & $0.10$ & $0.28$ & $0.4$ & $0.39$ & $0.9$ \\
        Efficiency map & $<0.01$ & $<0.01$ & $<0.01$ & $<0.01$ & $<0.01$ & $<0.01$ & $<0.01$\\
         Background in \Xicp\Km  & $ 0.02$ & $0.1$ & $<0.01$ & $<0.01$ & $0.04$ & $<0.01$ & $0.2$\\
         \hline
        Total & $0.9 $ & $ 2.5$ & $0.2 $ & $0.4$ & $ 0.5$ & $1.0 $ & $2.8$\\
        \hline
    \end{tabular}
    }
\end{table}

\clearpage

\clearpage

\newpage
\centerline
{\large\bf LHCb collaboration}
\begin
{flushleft}
\small
R.~Aaij$^{32}$,
C.~Abell{\'a}n~Beteta$^{50}$,
T.~Ackernley$^{60}$,
B.~Adeva$^{46}$,
M.~Adinolfi$^{54}$,
H.~Afsharnia$^{9}$,
C.A.~Aidala$^{86}$,
S.~Aiola$^{25}$,
Z.~Ajaltouni$^{9}$,
S.~Akar$^{65}$
% ...

{\footnotesize \it

% ...
$^{9}$Universit{\'e} Clermont Auvergne, CNRS/IN2P3, LPC, Clermont-Ferrand, France\\
% ...
$^{25}$INFN Sezione di Milano, Milano, Italy\\
% ...
$^{32}$Nikhef National Institute for Subatomic Physics, Amsterdam, Netherlands\\
% ...
$^{46}$Instituto Galego de F{\'\i}sica de Altas Enerx{\'\i}as (IGFAE), Universidade de Santiago de Compostela, Santiago de Compostela, Spain\\
% ...
$^{50}$Physik-Institut, Universit{\"a}t Z{\"u}rich, Z{\"u}rich, Switzerland\\
% ...
$^{54}$H.H. Wills Physics Laboratory, University of Bristol, Bristol, United Kingdom\\
% ...
$^{60}$Oliver Lodge Laboratory, University of Liverpool, Liverpool, United Kingdom\\
% ...
$^{65}$University of Cincinnati, Cincinnati, OH, United States\\
% ...
$^{68}$Syracuse University, Syracuse, NY, United States\\
% ...
$^{86}$University of Michigan, Ann Arbor, United States, associated to $^{68}$\\
\bigskip
% ...
\medskip
}
\end{flushleft}

\clearpage

\end{document}